\documentclass[reqno,11pt]{article}
\usepackage{psfig, amsmath, amstexnb, amsthm}

\newtheorem{theorem}{Theorem}[section]

\makeatletter

\def\enum{\ifnum \@enumdepth >3 \@toodeep\else
        \advance\@enumdepth \@ne 
        \edef\@enumctr{enum\romannumeral\the\@enumdepth}\list
        {\csname label\@enumctr\endcsname}
        {\setlength{\topsep}{1mm}
        \setlength{\parsep}{0mm}
        \setlength{\itemsep}{0mm}
        \setlength{\labelsep}{2mm}
        \settowidth{\leftmargin}{M.}
        \addtolength{\leftmargin}{\labelsep}
        \usecounter{\@enumctr}
        \def\makelabel##1{\hss\llap{##1}}}\fi}

\def\itemiz{\ifnum \@itemdepth >3 \@toodeep\else \advance\@itemdepth \@ne
        \edef\@itemitem{labelitem\romannumeral\the\@itemdepth}%
        \list{\csname\@itemitem\endcsname}{
        \setlength{\topsep}{1mm}
        \setlength{\parsep}{0mm}
        \setlength{\itemsep}{0mm}
        \setlength{\labelsep}{2mm}
        \settowidth{\leftmargin}{M.}
        \addtolength{\leftmargin}{\labelsep}
        \def\makelabel##1{\hss\llap{##1}}}\fi}

\def\captionheadfont@{\scshape}
\def\captionfont@{\small}
\long\def\@makecaption#1#2{%
  \setbox\@tempboxa\vbox{\color@setgroup
    \advance\hsize-3pc\noindent
    \captionfont@\captionheadfont@#1\@xp\@ifnotempty\@xp
        {\@cdr#2\@nil}{.\captionfont@\upshape\enspace#2}%
    \unskip\kern-3pc\par
    \global\setbox\@ne\lastbox\color@endgroup}%
  \ifhbox\@ne % the normal case
    \setbox\@ne\hbox{\unhbox\@ne\unskip\unskip\unpenalty\unkern}%
  \fi
  \ifdim\wd\@tempboxa=\z@ % this means caption will fit on one line
    \setbox\@ne\hbox to\columnwidth{\hss\kern-3pc\box\@ne\hss}%
  \else % tempboxa contained more than one line
    \setbox\@ne\vbox{\unvbox\@tempboxa\parskip\z@skip
        \noindent\unhbox\@ne\advance\hsize-3pc\par}%
\fi
  \ifnum\@tempcnta<64 % if the float IS a figure...
    \addvspace\abovecaptionskip
    \moveright 1.5pc\box\@ne
  \else % if the float IS NOT a figure...
    \moveright 1.5pc\box\@ne
    \nobreak
    \vskip\belowcaptionskip
  \fi
\relax
}

\makeatother

\DeclareMathSymbol{\leqsymb}{\mathalpha}{AMSa}{"36}
\def\leqs{\;\leqsymb\;}
\DeclareMathSymbol{\geqsymb}{\mathalpha}{AMSa}{"3E}
\def\geqs{\;\geqsymb\;}
\DeclareMathSymbol{\gtreqqlesssymb}{\mathalpha}{AMSa}{"54}

\newcommand{\field}[1]{\mathbb{#1}}

\newcommand{\Z}{\field{Z}\,}	% integers
\newcommand{\R}{\field{R}\,}	% real numbers
\newcommand{\cR}{{\mathcal R}}	% calligraphic N
\DeclareMathOperator{\e}{e}		% natural number
\DeclareMathOperator{\icx}{i}		% imaginary number
\DeclareMathOperator{\tg}{tan}		% tangent
\DeclareMathOperator{\Arctg}{atan}	% Arc tangent
\DeclareMathOperator{\dd}{d}		% integration measure
\DeclareMathOperator{\sign}{sign}	% sign
\def\vec#1{\ifmmode
\mathchoice{\mbox{\boldmath$\displaystyle\bf#1$}}
{\mbox{\boldmath$\textstyle\bf#1$}}
{\mbox{\boldmath$\scriptstyle\bf#1$}}
{\mbox{\boldmath$\scriptscriptstyle\bf#1$}}\else
{\mbox{\boldmath$\bf#1$}}\fi}

\def\math#1{\ifmmode
\mathchoice{\mbox{$\displaystyle\rm#1$}}
{\mbox{$\textstyle\rm#1$}}
{\mbox{$\scriptstyle\rm#1$}}
{\mbox{$\scriptscriptstyle\rm#1$}}\else
{\mbox{$\rm#1$}}\fi}		% roman style with adaptable size

\def\eps{\varepsilon}
\def\w{\omega}
\def\W{\Omega}
\def\x{\xi}
\def\y{\eta}
\def\z{\zeta}
\def\ph{\varphi}
\def\th{\theta}

\def\we{w_{\math e}}
\def\Je{J_{\math e}}
\def\wL{w_\Lambda}
\def\wK{w_K}

\def\defwd#1{{\sl #1}}			% defined word

\def\dx#1{\dd\!#1}			% integration measure
\def\dpar#1#2{\frac{\partial #1}{\partial #2}}	% partial derivative
\def\dtot#1#2{\frac{\dx{#1}}{\dx{#2}}}	% total derivative
\def\poisson#1#2{\{#1;#2\}}		% Poisson bracket

\def\brak#1{[#1]}			% bracket
\def\abs#1{\lvert#1\rvert}		% absolute value
\def\norm#1{\lVert#1\rVert}		% norm
\def\Order#1{{\mathcal O}(#1)}	% Order relation
\def\bigbrak#1{\bigl[#1\bigr]}		% big bracket
\def\bigpar#1{\bigl(#1\bigr)}		% big paranthesis
\def\Bigbrak#1{\Bigl[#1\Bigr]}		% Big bracket
\def\Bigpar#1{\Bigl(#1\Bigr)}		% Big paranthesis
\def\Bigevalat#1{\Bigr|_{#1}^{\phantom{#1}}}	% evalued at #1
\def\biggbrak#1{\biggl[#1\biggr]}	% bigg bracket
\def\avrg#1{\langle #1 \rangle}
\def\avvrg#1{\langle\langle #1 \rangle\rangle}

\def\figref#1{Fig.\ \ref{#1}}	% reference to a figure
\def\tabref#1{Table \ref{#1}}	% reference to a table
\def\writefig#1 #2 #3 {\rlap{\kern #1 truecm
\raise #2 truecm \hbox{\protect{\small #3}}}}

\def\bibtitle#1#2{#1, {\em #2}}                         % authors, title
\def\bibref#1#2#3#4#5{#1 {\bf #2}:#3--#4 (#5)}        % review, year
\def\bibarticle#1#2#3#4#5#6#7{\bibtitle{#1}{#2},
\bibref{#3}{#4}{#5}{#6}{#7}.}

\def\bibbook#1#2#3#4{#1, {\em #2} (#3, #4).}

\def\PR{Phys.\ Rev.}
\def\PRA{Phys.\ Rev.\ A}
\def\PRB{Phys.\ Rev.\ B}
\def\PRE{Phys.\ Rev.\ E}
\def\PRL{Phys.\ Rev.\ Lett.}
\def\RMP{Rev.\ Mod.\ Phys.}

\begin{document}

%%%%%%%%%%%%%%%%%%%%%%%%%%%%%%%%%%%%%%%%%%%%%%%%%%%%%%%%%%%%%%%%%%%%%%%%%%%%%%%%%

\title{The averaged dynamics of the hydrogen atom\\
in crossed electric and magnetic fields\\
as a perturbed Kepler problem\thanks{Dedicated to Martin C.\ Gutzwiller on
the occasion of his 75th birthday.}}
\author{Nils Berglund\thanks{Present address: Weierstra\ss\ Institut,
Mohrenstra\ss e 39, D-10117 Berlin, Germany}  and Turgay Uzer\\
School of Physics, 
Georgia Tech\\
Atlanta, GA 30332-0430, USA\\
%{\tt ph297nb@prism.gatech.edu}
}
\date{July 13, 2000}
\maketitle

\begin{abstract}
We treat the classical dynamics of the hydrogen atom in perpendicular
electric and magnetic fields as a celestial mechanics problem. By
expressing the Hamiltonian in appropriate action--angle variables, we
separate the different time scales of the motion. The method of averaging
then allows us to reduce the system to two degrees of freedom, and to
classify the most important periodic orbits. 
\end{abstract}

%%%%%%%%%%%%%%%%%%%%%%%%%%%%%%%%%%%%%%%%%%%%%%%%%%%%%%%%%%%%%%%%%%%%%%%%%%%%%%%%%

\section{Introduction}
\label{sec_in}

Our contribution to this Special Issue lies at the intersection of Martin
Gutzwiller's scientific interests, namely celestial mechanics, electron
motion, and chaos, especially its manifestations in quantal systems. We
will be performing ``celestial mechanics on a microscopic scale''
\cite{UDMRS91} by treating the dynamics of highly excited (``Rydberg'')
electrons \cite{Connerade} in crossed electric and magnetic fields using
classical mechanics.

Rydberg atoms in strong external fields constitute fundamental physical
systems where the quantum mechanical regime of strong nonlinearity can be
tested \cite{Gutzwiller,Kock/vanLeeuwen}. While the problem of a Rydberg
atom interacting with a strong magnetic field (the Diamagnetic Kepler
Problem, DKP, also known as the Quadratic Zeeman Effect, QZE) has been
fairly well understood as a result of sustained research in the past two
decades \cite{Hasegawa/Robnik/Wunner,Friedrich/Wintgen}, the superficially
similar scenario resulting from the addition of a perpendicular electric
field -- the so-called crossed field arrangement
\cite{Solovev,Braun/Solovev,Wiebusch,Raithel/Fauth/Walther, 
vonMilczewski/Diercksen/Uzer,vonMilczewski/Diercksen/Uzer2} -- remains the
least understood of all Rydberg problems. This is all the more remarkable
in view of the prominence of the crossed fields in diverse areas of physics
ranging from excitonic systems to plasmas and neutron stars. This problem
is so complex because no continuous symmetry survives the extensive
symmetry breaking \cite{Delande/Gay} induced by the two fields. The result
is a wealth of new physics which is only possible beyond two degrees of
freedom, such as Arnol'd diffusion
\cite{TLL79,Gutzwiller,Lichtenberg/Lieberman,vonMilczewski/Diercksen/Uzer3}.
This absence of symmetry also allows localizing electronic wavepackets in
all spatial dimensions, and the observation of these localized wavepackets
\cite{Yeazell} has led to new insights into the dynamics of the electron in
the correspondence principle regime. It has also been found that a
velocity-dependent, Coriolis-like force in Newton's equations causes the
ionization of the electron to exhibit chaotic scattering
\cite{Main/Wunner2,Uzer/Farrelly2,JFU99}. All these phenomena, as well as
renewed interest in the motional Stark effect
\cite{Johnson/Hirschfelder/Yang,Farrelly}, make the crossed-fields problem
an experimental accessible paradigm for a wide variety of outstanding
issues in atomic and molecular physics, solid-state physics
\cite{Digman/Sipe,Schmelcher}, nuclear physics \cite{Bohr/Mottelson},
astrophysics \cite{Mathys}, and celestial mechanics \cite{Mignard}.

The experimental challenge has been taken up by Raithel, Fauth, and
Walther  \cite{Raithel/Fauth/Walther,Raithel/Fauth/Walther2} who in a
landmark series of experiments have identified a class of quasi-Landau (QL)
resonances in the spectra of rubidium Rydberg atoms in crossed electric and
magnetic fields. Similar to the original QL resonances observed by Garton
and Tomkins \cite{Garton/Tomkins}, this set of resonances is associated
with a rather small set of {\em planar} orbits of the crossed-fields
Hamiltonian which is known to support an enormous number of mostly
non-planar periodic motions \cite{Raithel/Fauth/Walther}. The dominance of
planar orbits in these experiments has recently been explained
\cite{vMFU97}.

In contrast with the DKP, and despite some preliminary work \cite{FW96},
the systematics of periodic orbits in the crossed-fields problem has not
been discovered up to now. The aim of the present work is to initiate a
systematic classification of the orbits of the crossed-fields Hamiltonian,
based on methods developed in celestial mechanics, specifically Delaunay
variables and averaging. The analogy between atomic and planetary systems
was already used by Born \cite{Born}, who studied in particular the
crossed-fields problem, but neglected the quadratic Zeeman term because he
was not studying Rydberg atoms, where it is prominent.

Delaunay variables are action--angle variables which have a clear geometric
interpretation in terms of Kepler ellipses. A fascinating historical
account of the developments in celestial mechanics leading to the
introduction of Delaunay variables can be found in a recent review by
Gutzwiller \cite{Gutzwiller2}. These variables allow to separate the time
scales of the motion, which is represented as a fast rotation of the planet
(or the electron), along a Kepler ellipse with slowly changing orientation
and eccentricity. 

The technique of averaging \cite{V96} allows to decrease the number of
degrees of freedom by eliminating the fast motion along the Kepler ellipse.
It has been used for a long time in celestial mechanics to compute the
so-called secular motion of the solar system, and can be considered as a
first order perturbation theory \cite{Deprit,Henrard}. A systematic use of
averaging allowed Laskar to integrate the motion of the solar system over
several hundred million years \cite{Laskar1,Laskar2,LR93}. The method has
been applied to the DKP in \cite{DKN83,CDMW87}. 

We note in passing that  the direct connection to celestial mechanics that
the Delaunay variables provide is lost with an alternative set of variables
called the Lissajous elements. These are obtained by regularizing the
Coulomb Hamiltonian \cite{Farrelly/Uzer/&92}, and are appropriate for
investigating the level structure of Rydberg atoms. The connection between
the two sets in two dimensions is given by \cite{DW91}.

In this paper, we use the following notations. 
The Hamiltonian of an electron subjected to a Coulomb potential, a magnetic
field $\vec B$ and an electric field $\vec F$ can be written in
dimensionless units as
\begin{equation}
\label{i1}
H(p_x,p_y,p_z;x,y,z) = \frac12 p^2 - \frac1r + \frac12 \vec L\cdot\vec B +
\frac18(\vec r\wedge\vec B)^2 - \vec r\cdot\vec F, 
\end{equation}
where $\vec r = (x,y,z)$ is the electron's position, $\vec p =
(p_x,p_y,p_z)$ its momentum, and $\vec L = \vec r \wedge \vec p$ its angular
momentum. We write $r=\norm{\vec r}$ and $p=\norm{\vec p}$.

Our paper is organized as follows. In Section \ref{sec_ze}, we summarize
previous results on the case when only the magnetic field is present. 
This allows us to introduce Delaunay variables and the method of averaging,
and illustrate them in a relatively simple situation.

In Section \ref{sec_s}, we consider the case when only an electric field is
present, which is integrable \cite{R63}. We introduce another set of
action--angle variables (``electric action--angle variables'') based on
parabolic variables, which are better adapted to perturbation theory
\cite{Born}. We then derive (new) transformation formulas from electric
action--angle variables to the geometrically more transparent Delaunay
variables.

With these tools and sets of coordinates in mind, we finally turn to the
crossed-fields Hamiltonian \eqref{i1} in Section \ref{sec_cr}. We start by
considering the two limiting cases $B\ll F$ and $F\ll B$, which involve
three distinct time scales, and can thus be analysed by a second averaging.
We then study the dynamics in the plane perpendicular to $\vec B$ for
general values of the fields. We conclude by an overview of the general
structure of the phase space of the averaged Hamiltonian.

%%%%%%%%%%%%%%%%%%%%%%%%%%%%%%%%%%%%%%%%%%%%%%%%%%%%%%%%%%%%%%%%%%%%%%%%%%%%%%%%%

\section{The Quadratic Zeeman Effect (or DKP)}
\label{sec_ze}

We start by considering Hamiltonian \eqref{i1} in the case $F=0$. If we take
the $z$-axis along the magnetic field $\vec B$, \eqref{i1} can be written as
\begin{equation}
\label{z1}
H = \frac12 p^2 - \frac1r + \frac B2 L_z + \frac{B^2}8 (x^2+y^2).
\end{equation}
Although the equations of motion are easily written down, it is difficult to
understand the qualitative properties of dynamics in cartesian coordinates.
We will therefore take advantage of the fact that for small $B$, \eqref{z1}
is a small perturbation of the integrable Kepler problem, for which
action--angle variables are known explicitly. By writing the Hamiltonian
\eqref{z1} in these variables, we can separate slow and fast
components of the motion. The qualitative dynamics can then be further
analysed by using the method of averaging.

%%%%%%%%%%%%%%%%%%%%%%%%%%%%%%%%%%%%%%%%%%%%%%%%%%%%%%%%%%%%%%%%%%%%%%%%%%%%%%%%%

\subsection{Delaunay variables}
\label{sec_zdv}

We start by considering the Kepler Hamiltonian
\begin{equation}
\label{zd1}
H = \frac12 p^2 - \frac1r.
\end{equation}
Besides the energy $H$, it admits as constants of motion the angular
momentum $\vec L$ and the Runge-Lenz vector
\begin{equation}
\label{zd2}
\vec A = \frac{\vec r}{r} + \vec L \wedge \vec p.
\end{equation}
If $H<0$, the motion takes place in a plane perpendicular to $\vec L$, on
an ellipse of eccentricity $e = \norm{\vec A}$ and major axis parallel to
$\vec A$ and of length $2a = -1/H$. 

Action--angle variables taking these properties into account are well known
in celestial mechanics, where they are called \defwd{Delaunay variables}.
The action variables are given by\footnote{We use the letters $G$ and $K$ in
order to distinguish the action variables from their physical meaning. In
celestial mechanics, one usually denotes $L_z$ by $H$ instead of $K$, but we
prefer the latter notation in order to avoid confusion with the Hamiltonian.}
\begin{equation}
\label{zd3}
\begin{split}
\Lambda &= \sqrt{a} \\
G &= \sqrt{a} \sqrt{1-e^2} = \norm{\vec L} \\
K &= \sqrt{a} \sqrt{1-e^2} \,\cos i = L_z,
\end{split}
\end{equation}
where the inclination $i\in[0,\pi]$ is the angle between $\vec L$ and the
$z$-axis. These variables are defined on the domain $\abs{K}\leqs G\leqs
\Lambda$. 

The corresponding angle variables are defined in the following way. The
intersection between the plane of the orbit and the $xy$-plane is called
\defwd{line of nodes}. The angle $\W$ between line of nodes and $x$-axis is
called \defwd{longitude of nodes} and is conjugated to $K$; the angle $\w$
between major axis of the Kepler ellipse and line of nodes is called
\defwd{argument of perihelion} and is conjugated to $G$; the \defwd{mean
anomaly} $M$, which is conjugated to $\Lambda$, is proportional to the area
swept on the ellipse, according to Kepler's second law.

\begin{figure}
 \centerline{\psfig{figure=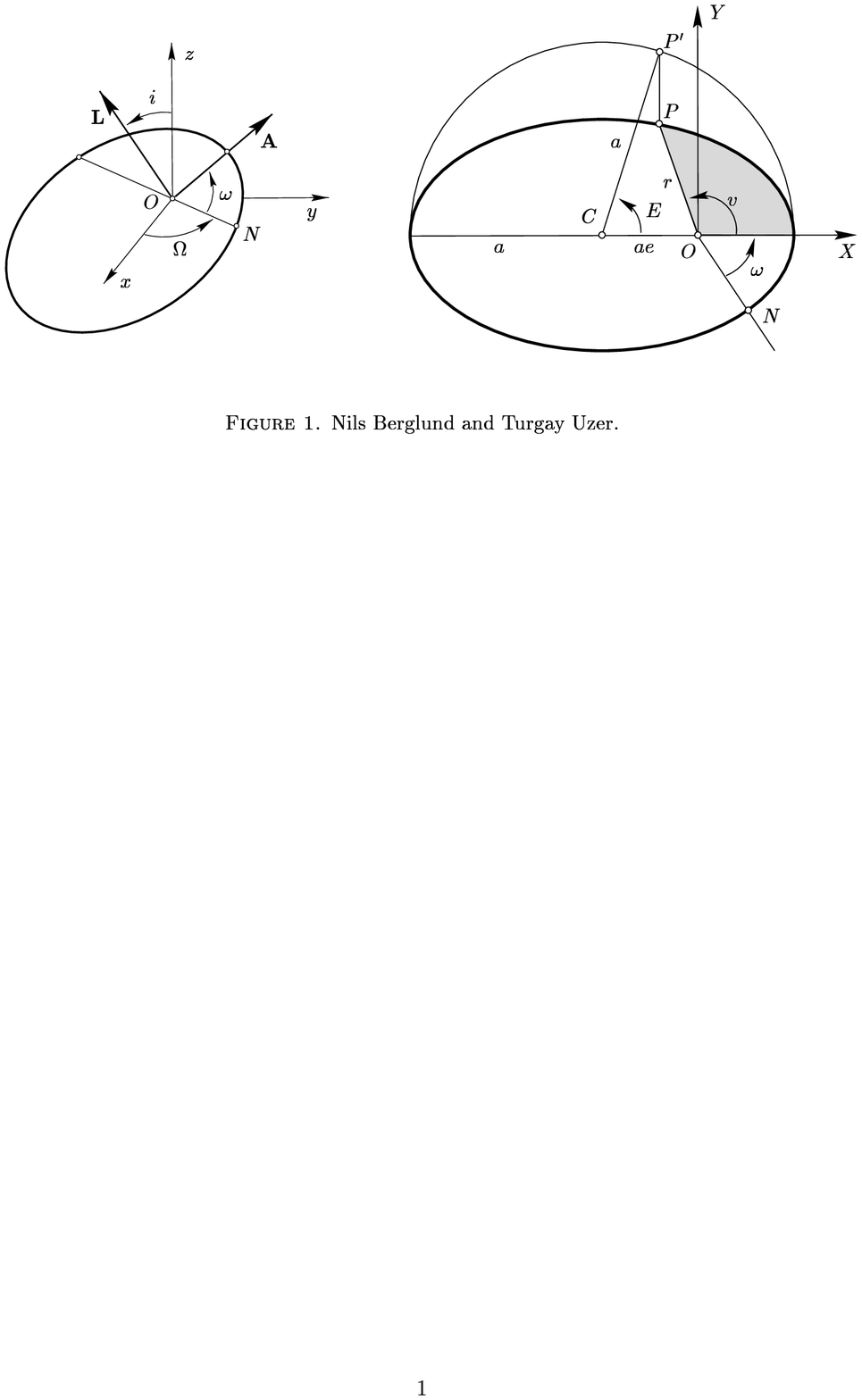,height=60mm,clip=t}}
 \vspace{1mm}
 \caption[]
 {Definition of Delaunay variables. The angles $\W$, $i$ and $\w$ determine
 the position of the Kepler ellipse in space. The line of nodes $ON$ is the
 intersection of the plane of the ellipse with the $xy$-plane. The shaded
 area is proportional to the mean anomaly $M$. The true anomaly $v$ and
 eccentric anomaly $E$ are auxiliary quantities, allowing to relate the
 $(X,Y)$-coordinates of the planet $P$ with $M$.}
\label{fig_z1}
\end{figure}

In order to compute $M$, we introduce orthogonal coordinates $(X,Y)$, where
$X$ is attached to the major axis of the ellipse. The \defwd{true anomaly}
$v$ and the \defwd{eccentric anomaly} $E$ are auxiliary quantities defined
by the relations
\begin{equation}
\label{zd4}
\begin{split}
X &= r\cos v = a (\cos E-e) \\
Y &= r\sin v = a \sqrt{1-e^2} \sin E,
\end{split}
\end{equation}
see \figref{fig_z1}. By trigonometry one can show that $M$ and $E$ are related by 
\defwd{Kepler's equation}
\begin{equation}
\label{zd5}
M = E - e \sin E,
\end{equation}
which implies in particular that
\begin{equation}
\label{zd6}
\dtot ME = 1 - e\cos E = \frac ra .
\end{equation}
The transition from Delaunay variables to cartesian coordinates is done in the
following way. Given the actions \eqref{zd3}, we can compute
\begin{equation}
\label{zd7}
a = \Lambda^2, \qquad 
e^2 = 1 - \frac{G^2}{\Lambda^2}, \qquad
\cos i = \frac KG.
\end{equation}
We can then determine $X$, $Y$ and the conjugated momenta
\begin{equation}
\label{zd8}
\begin{split}
P_X &= -\frac1{\sqrt a} \frac{\sin E}{1-e\cos E} \\
P_Y &= \frac1{\sqrt a} \frac{\sqrt{1-e^2}\cos E}{1-e\cos E}.
\end{split}
\end{equation}
The cartesian coordinates are then given by the relations 
\begin{equation}
\label{zd9}
\begin{pmatrix}
x & p_x \\ y & p_y \\ z & p_z
\end{pmatrix}
= \cR_z(\W) \cR_x(i) \cR_z(\w) 
\begin{pmatrix}
X & P_X \\ Y & P_Y \\ 0 & 0
\end{pmatrix},
\end{equation}
where $\cR_x$ and $\cR_z$ describe rotations around the $x$- and $z$-axis,
given by
\begin{equation}
\label{zd10}
\cR_x(i) = 
\begin{pmatrix}
1 & 0 & 0 \\ 
0 & \cos i & -\sin i \\ 
0 & \sin i & \phantom{-}\cos i
\end{pmatrix},
\qquad
\cR_z(\w) = 
\begin{pmatrix}
\cos \w & -\sin \w & 0\\ 
\sin \w & \phantom{-}\cos \w & 0\\
0 & 0 & 1
\end{pmatrix}.
\end{equation}
It is then straightforward to show that the Hamiltonian \eqref{zd1} takes
the form 
\begin{equation}
\label{zd11}
H = -\frac1{2\Lambda^2},
\end{equation}
and thus the equations of motion are given by
\begin{align}
\nonumber
\dot{\Lambda} &= 0 	& \dot{M} &= \frac{1}{\Lambda^3} \\
\label{zd12}
\dot{G} &= 0 	& \dot{\w} &= 0 \\
\nonumber
\dot{K} &= 0 	& \dot{\W} &= 0.
\end{align}
Besides the actions $\Lambda$, $G$, $K$, the two angles $\w$ and $\W$ are
also constants, which reflects the high degeneracy of the hydrogen atom.
Equation \eqref{zd11} also gives a physical interpretation of $\Lambda$ as a
function of the energy. In quantum mechanics, $\Lambda$ corresponds to
the principal quantum number.

Let us now return to the Zeeman effect. The Hamiltonian \eqref{z1} can be
expressed in Delaunay variables as
\begin{equation}
\label{zd13}
\begin{split}
H &= -\frac1{2\Lambda^2} + \frac B2 K + B^2 H_1(\Lambda,G,K;M,\w) \\
H_1 &= \tfrac1{16} r^2\bigbrak{1 + \cos^2 i + \sin^2 i \cos(2\w+2v)},
\end{split}
\end{equation}
where $r$, $\sin i$ and $v$ can be expressed in terms of Delaunay variables
using \eqref{zd4}, \eqref{zd6} and \eqref{zd7}. The equations of motion take
the form
\begin{align}
\nonumber
\dot{\Lambda} &= B^2 \poisson{\Lambda}{H_1}  & 
\dot{M} &= \frac1{\Lambda^3} + B^2 \poisson{M}{H_1} \\
\label{zd14}
\dot{G} &= B^2 \poisson{G}{H_1}  & 
\dot{\w} &= \phantom{\frac1{\Lambda^3} +{}} B^2 \poisson{\w}{H_1} \\
\nonumber
\dot{K} &=0 & 
\dot{\W} &= \frac B2 + B^2 \poisson{\W}{H_1}, 
\end{align}
where the Poisson bracket is defined by
\begin{equation}
\label{zd15}
\poisson fg = 
\dpar fM \dpar g\Lambda - \dpar f\Lambda \dpar GM 
+ \dpar f\w \dpar gG - \dpar fG \dpar g\w 
+ \dpar f\W \dpar gK - \dpar fK \dpar g\W.
\end{equation}
We discuss the computation of these Poisson brackets in Appendix
\ref{app_pb} (see in particular \tabref{t_z1}).

To first order in $B$, \eqref{zd14} describes the Larmor precession of the
ellipse. Since the Hamiltonian does not depend on $\W$, $K=L_z$ is a
constant of the motion and \eqref{zd14} is in effect a
two-degrees-of-freedom system. For small $B$, $M$ is a fast variable, while
$\Lambda$, $G$ and $\w$ are slow ones. The motion can thus be imagined as a
fast motion of the electron along a slowly ``breathing'' and rotating
ellipse. The dynamics can be visualized by a Poincar\'e map, taking for
instance a section at constant $M$, and plotting the value of $G$ and $\w$
at each intersection \cite{DKN83}.

%%%%%%%%%%%%%%%%%%%%%%%%%%%%%%%%%%%%%%%%%%%%%%%%%%%%%%%%%%%%%%%%%%%%%%%%%%%%%%%%%

\subsection{Averaging}
\label{ssec_za}

To analyse the motion of \eqref{zd14} for small $B$, one can use the fact
that $M$ is the only fast variable, so that the dynamics of the slow
variables will be essentially determined by the average effect of $M$ during
one period. 

With any given function $f$ of the Delaunay variables, let us associate its 
average
\begin{equation}
\label{za1}
\avrg{f}_\Lambda(G,K;\w,\W) = \frac1{2\pi} \int_0^{2\pi}
f(\Lambda,G,K;M,\w,\W) \dx M.
\end{equation}
The averaged Hamiltonian $\avrg{H}_\Lambda$ generates the
canonical equations
\begin{align}
\nonumber
\dot{\Lambda} &= 0  & 
\dot{M} &= \frac1{\Lambda^3} + B^2 \poisson{M}{\avrg{H_1}_\Lambda} \\
\label{za2}
\dot{G} &= B^2 \poisson{G}{\avrg{H_1}_\Lambda}  & 
\dot{\w} &= \phantom{\frac1{\Lambda^3} +{}} B^2
\poisson{\w}{\avrg{H_1}_\Lambda} \\
\nonumber
\dot{K} &=0 & 
\dot{\W} &= \frac B2 + B^2 \poisson{\W}{\avrg{H_1}_\Lambda}.
\end{align}
Since $\avrg{H_1}_\Lambda$ does not depend on $M$ and $\W$,
$\avrg{H}_\Lambda$ is in effect a one-degree-of-freedom Hamiltonian,
depending on $K$ and $\Lambda$ as on parameters.

A standard result from averaging theory (see Appendix \ref{app_av}) states
that the equations \eqref{za2} are a good approximation of the equations
\eqref{zd14}, in the sense that
\begin{itemiz}
\item	orbits of \eqref{za2} and \eqref{zd14} with the same initial
condition differ by a term of order $B^2$ during a time of order $1/B^2$;
\item	to each nondegenerate equilibrium of \eqref{za2}, there corresponds
a periodic orbit of \eqref{zd14}, at a distance of order $B^2$ of the
equilibrium, and which has the same stability if the equilibrium is
hyperbolic.
\end{itemiz}

\begin{table}
\begin{center}
\begin{tabular}{|c|c||c|c|}
\hline
\vrule height 12pt depth 8pt width 0pt
$f$ & $\avrg{f}_\Lambda$ & $f$ & $\avrg{f}_\Lambda$ \\
\hline
\vrule height 14pt depth 8pt width 0pt
$X$ & $-\frac32 ae$ & 
$X^2$ & $a^2(\frac12+2 e^2)$ \\
\vrule height 12pt depth 8pt width 0pt
$Y$ & $0$ & 
$Y^2$ & $a^2 (\frac12 - \frac12 e^2)$ \\
\vrule height 12pt depth 8pt width 0pt
$r$ & $a(1+\frac12 e^2)$ & 
$r^2$ & $a^2(1+\frac32 e^2)$ \\
\vrule height 12pt depth 9pt width 0pt
$z$ & $-\frac32 a e \sin i \sin \w$ & 
$z^2$ & $a^2 \sin^2 i \bigbrak{\frac12 + \frac14 e^2 (3-5\cos 2\w)}$\\
\hline
\end{tabular}
\end{center}
\caption[]
{Some important quantities and their averages over the fast variable $M$.}
\label{t_z2}
\end{table}

Averages over $M$ can be computed quite easily by the formula
\begin{equation}
\label{za3}
\avrg{f}_\Lambda = \frac1{2\pi} \int_0^{2\pi} f(M) \dtot ME \dx E 
= \frac1{2\pi} \int_0^{2\pi} f(M(E)) (1-e\cos E)\dx E.
\end{equation}
Some useful averages are given in \tabref{t_z2}. The averaged Hamiltonian
can be written in the form \cite{CDMW87}
\begin{equation}
\label{za4}
\avrg{H_1}_\Lambda = \tfrac1{16} \Lambda^4 \Bigbrak{(1+\cos^2 i)(1+\tfrac32
e^2) + \tfrac52 e^2 \sin^2i \cos 2\w},
\end{equation}
from which we deduce the relevant equations of motion
\begin{equation}
\label{za5}
\begin{split}
\dot{G} &= \tfrac5{16} B^2 \Lambda^4 e^2 \sin i \sin 2\w \\
\dot{\w} &= \tfrac1{16} B^2 \frac{\Lambda^4}G \Bigbrak{3(e^2-1) - 5\cos^2 i +
5(e^2-\sin^2i)\cos 2\w}.
\end{split}
\end{equation}
The system can now be studied by analysing the orbits of \eqref{za5} or,
equivalently, the level lines of the function \eqref{za4} in the
$(\w,G)$-plane \cite{DKN83}. Doing this, one finds that the vector field
has two elliptic equilibrium points located at
\begin{equation}
\label{za6}
\w = \frac{\pi}2, \frac{3\pi}2, \qquad
G^2 = \sqrt5 \abs{K}\Lambda,
\end{equation}
which exist if $\abs{K}<\Lambda/\sqrt5$. 

The vector field behaves in a singular way at the boundaries $G=\abs{K}$
and $G=\Lambda$ of the domain. These singularities have been explained by
Coffey and coworkers \cite{CDMW87}, who showed that the phase space has the
topology of a sphere. The line $G=\Lambda$, $\w\in[0,2\pi)$ has to be
contracted into the north pole of the sphere, corresponding to $e=0$ and
thus to circular motion; indeed, in this case the perihelion, and hence its
argument $\w$, is undefined. The line $G=\abs{K}$, $\w\in[0,2\pi)$ has to be
contracted into the south pole of the sphere, corresponding to $i=0$ and
thus to equatorial motion; in that case, the sum $\W+\w$ is sufficient to
specify the position of the ellipse.

\begin{figure}
 \centerline{\psfig{figure=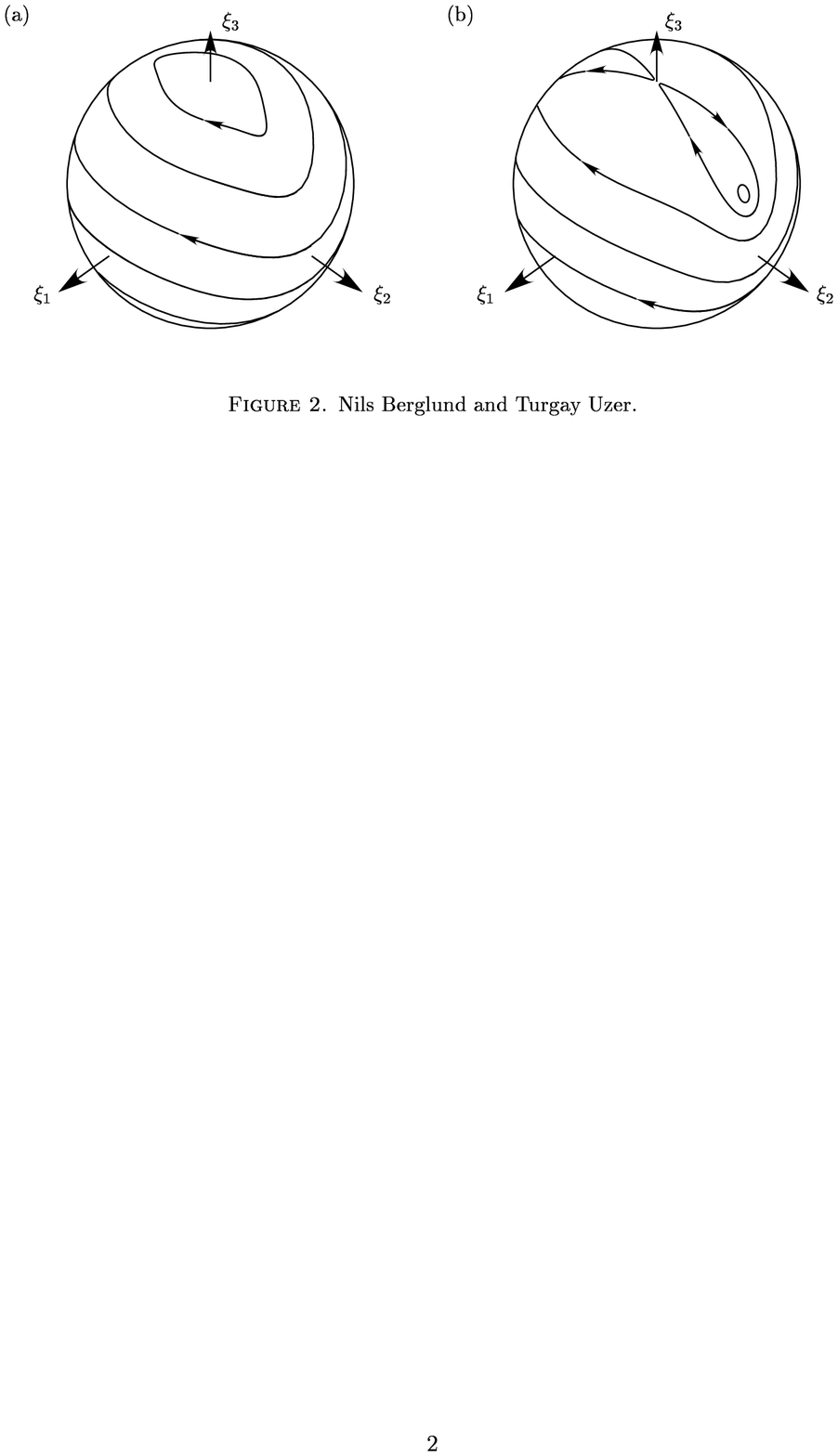,height=55mm,clip=t}}
 \vspace{1mm}
 \caption[]
 {Phase portraits of the averaged Hamiltonian \eqref{za3} on the sphere
 $\x_1^2 + \x_2^2 + \x_3^2 = \bigpar{\frac12 (\Lambda^2-K^2)}^2$, (a) for
 $\abs{K} > \Lambda/\sqrt5$ and (b) for $\abs{K} < \Lambda/\sqrt5$.  The
 north pole corresponds to circular C-orbits, the south pole (not shown) to
 equatorial B-orbits, which are elliptic in both cases. In the second case,
 two additional periodic orbits (the Z-orbits) appear in the plane
 $\x_1=0$.}
\label{fig_z2}
\end{figure}

To account for the spherical topology of phase space, \cite{CDMW87} have
introduced the variables
\begin{equation}
\label{za7}
\begin{split}
\x_1 &= G\Lambda e \sin i \,\cos \w \\
\x_2 &= G\Lambda e \sin i \,\sin \w \\
\x_3 &= G^2 - \tfrac12(\Lambda^2+K^2),
\end{split}
\end{equation}
which belong to the sphere
\begin{equation}
\label{za8}
\x_1^2 + \x_2^2 + \x_3^2 = \Bigpar{\frac{\Lambda^2-K^2}2}^2.
\end{equation}
Using the fact that $\avrg{H_1}_\Lambda$ can be put into the form
\begin{equation}
\label{za9}
\avrg{H_1}_\Lambda = \frac{\Lambda^4}{16} \Bigbrak{1+\cos^2 i + 3 e^2 +
\frac{\x_1^2-4\x_2^2}{\Lambda^2 G^2}},
\end{equation}
and the Poisson brackets in Table \ref{t_z3} of Appendix \ref{app_pb}, 
they derive the equations of motion 
\begin{equation}
\label{za10}
\begin{split}
\dot{\x_1} &= \frac{\Lambda^4}{8G} \,\x_2 \Bigbrak{\frac{5\x_1^2}{\Lambda^2
G^2} - 1 + e^2 + 5 \cos^2 i} \\
\dot{\x_2} &= \frac{\Lambda^4}{8G} \,\x_1 \Bigbrak{\frac{5\x_2^2}{\Lambda^2
G^2} - 4(1-e^2)} \\
\dot{\x_3} &= \frac{5\Lambda^2}{4G} \,\x_1\x_2.
\end{split}
\end{equation}
In these variables, the poles have become equilibrium points around which
the flow is nonsingular, so that their stability can be easily determined. 
Depending on the relative value of the constants of motion $\Lambda$ and
$K$, there are two qualitatively different phase portraits
(\figref{fig_z2}): 

\begin{enum}
\item	If $\abs{K} > \Lambda/\sqrt5$, both poles are elliptic, and there
are no other equilibrium points. All other orbits rotate around the sphere
with $\dot\w<0$ (\figref{fig_z2}a).

\item	If $\abs{K} < \Lambda/\sqrt5$, the south pole is still elliptic, but
the north pole has become hyperbolic, and the new elliptic equilibria
\eqref{za6} have appeared in a pitchfork bifurcation. Two homoclinic
orbits of the north pole separate orbits rotating around each of the three
elliptic equilibria (\figref{fig_z2}b).
\end{enum}

The averaging theorem shows that to each of the four possible equilibrium
points of \eqref{za10}, there corresponds a periodic orbit of the exact
system \eqref{zd14} (see also Appendix \ref{app_pb}). For further
reference, let us call B-orbits the equatorial orbits (which lie in the
plane perpendicular to $\vec B$), C-orbits the circular ones, and Z-orbits
those corresponding to the nontrivial equilibrium \eqref{za6}.  One further
expects that the periodic orbits of \eqref{za10} approximate either
quasiperiodic KAM-type orbits of \eqref{zd14}, or ``soft'' chaotic
components associated with resonances. More prominent chaotic components
are expected near the homoclinic orbits of the north pole.

In other works \cite{Hasegawa/Robnik/Wunner}, orbits are sometimes
represented in cylindric coordinates $(\rho,\phi,z)$. They can be deduced
from Delaunay variables by the relations 
\begin{equation}
\label{za11}
\begin{split}
z &= r \sin i \sin(\w+v) \\
\rho &= r \sqrt{1-\sin^2i\sin^2(\w+v)}.
\end{split}
\end{equation}
The different periodic orbits can thus be parametrized either by the true
anomaly $v$ or by the eccentric anomaly $E$ as
\begin{align}
\nonumber
&\text{C-orbits} &
z &= \Lambda\sqrt{\Lambda^2-K^2}\,\sin v &
\rho &= \Lambda\sqrt{\Lambda^2\cos^2 v + K^2\sin^2 v} \\
\label{za12}
&\text{B-orbits} &
z &= 0 &
\rho &= \Lambda (\Lambda-\sqrt{\Lambda^2-K^2}\cos E) \\
\nonumber
&\text{Z-orbits} &
z &= \Lambda^2 \sin i \,(\cos E-e) &
\rho &= \Lambda^2 \sqrt{(1-e\cos E)^2 - \sin^2i\,(\cos E-e)^2},
\end{align}
where $\sin^2i = 1-\abs{K}/\sqrt5 \Lambda$ and $e^2 =
1-\sqrt5\abs{K}/\Lambda$ in the last case. 
The C-, B- and Z-orbits are labelled, respectively, $C$, $I_1$ and
$I_\infty$ in \cite{Hasegawa/Robnik/Wunner}.

%%%%%%%%%%%%%%%%%%%%%%%%%%%%%%%%%%%%%%%%%%%%%%%%%%%%%%%%%%%%%%%%%%%%%%%%%%%%%%%%%

%\newpage
\section{The Stark effect}
\label{sec_s}

We consider now the Hamiltonian \eqref{i1} in the case $B=0$. If we choose
the $z$-axis along the electric field $\vec F$, it can be written as 
\begin{equation}
\label{s1}
H = \frac12 p^2 - \frac1r + Fz.
\end{equation}
Besides the energy and the $z$-component of the angular momentum, this
system has a third constant of the motion and is thus integrable. Indeed, as
shown by \cite{R63}, the generalization of the Runge-Lenz vector
\eqref{zd2},
\begin{equation}
\label{s2}
\vec C = \vec A - \frac12 \bigpar{\vec r\wedge \vec F} \wedge \vec r
\end{equation}
satisfies the equation of motion
\begin{equation}
\label{s3}
\dot{\vec C} = \frac32 \vec L \wedge \vec F,
\end{equation}
and thus $C_z = \vec C \cdot \vec F$ is constant.

We will start, in Section \ref{ssec_sd}, by analysing the system in
Delaunay variables, in particular in its averaged form, in order to get a
feeling for the geometry of the orbits. In Section \ref{ssec_saa}, we
present another description of the system, based on parabolic variables,
which allows to construct action--angle variables taking the constant $C_z$
into account. Though they are better suited for perturbation theory, these
action--angle variables have a less obvious geometric interpretation than
Delaunay variables. This is why we establish the transformation formulas
between both sets of variables in Section \ref{ssec_sc}, in the limit $F\to
0$.

%%%%%%%%%%%%%%%%%%%%%%%%%%%%%%%%%%%%%%%%%%%%%%%%%%%%%%%%%%%%%%%%%%%%%%%%%%%%%%%%%

\subsection{Delaunay variables}
\label{ssec_sd}

\begin{figure}
 \centerline{\psfig{figure=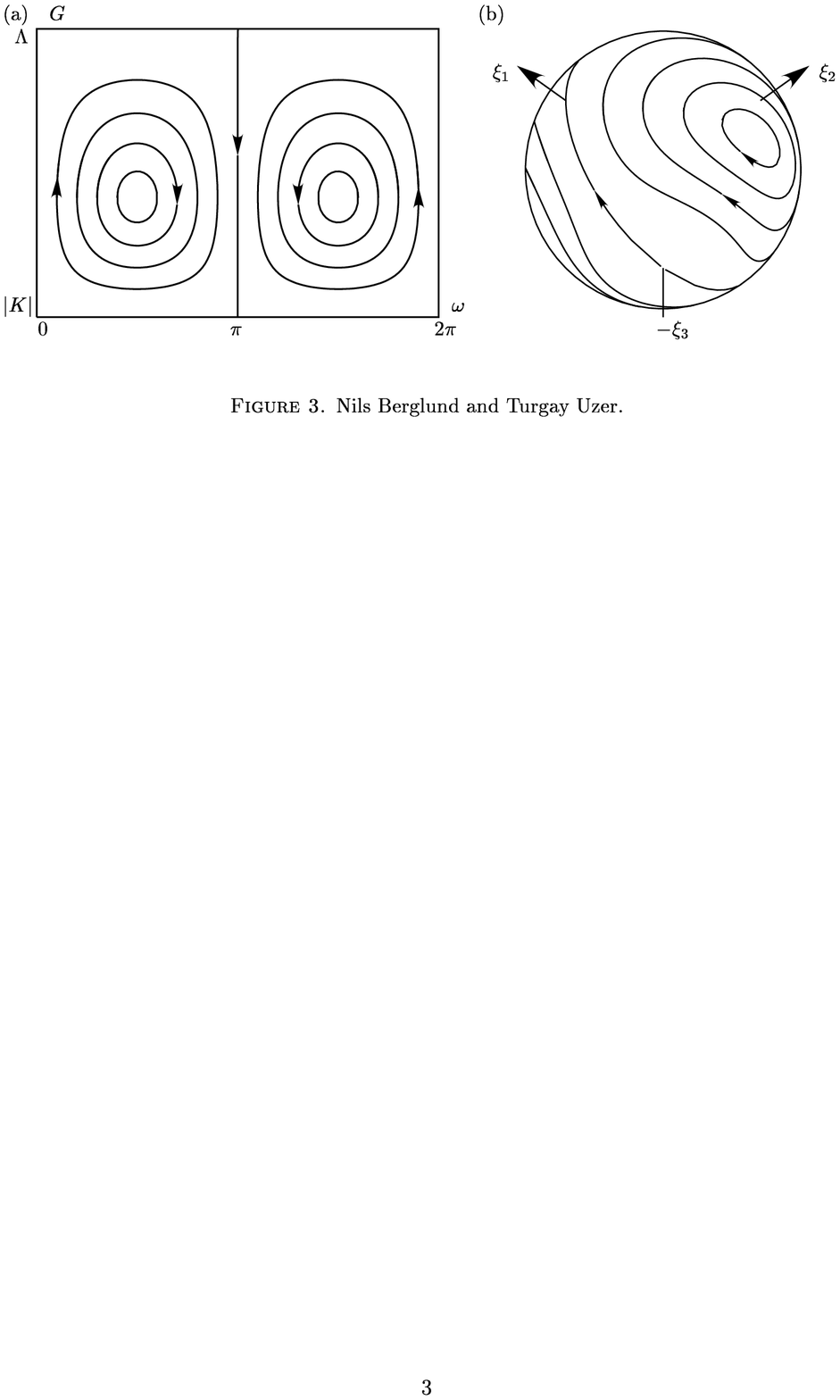,height=55mm,clip=t}}
 \vspace{1mm}
 \caption[]
 {Phase portraits of the averaged Hamiltonian \eqref{sd4} (a) in the
 $(\w,G)$-plane and (b) on the sphere $\x_1^2 + \x_2^2 + \x_3^2 =
 \bigpar{\frac12 (\Lambda^2-K^2)}^2$, seen from below the south pole. 
 The elliptic points at $\w =\frac\pi2$, $\frac{3\pi}2$ correspond to
 periodic S-orbits of the Stark Hamiltonian.}
\label{fig_s1}
\end{figure}

In Delaunay variables, the Hamiltonian \eqref{s1} takes the form 
\begin{equation}
\label{sd1}
\begin{split}
H &= -\frac1{2\Lambda^2} + F H_2(\Lambda,G,K;M,\w), \\
H_2 &= r \sin i\,\sin(\w+v),
\end{split}
\end{equation}
and the equations of motion have the structure
\begin{align}
\nonumber
\dot{\Lambda} &= F \poisson{\Lambda}{H_2}  & 
\dot{M} &= \frac1{\Lambda^3} + F \poisson{M}{H_2} \\
\label{sd2}
\dot{G} &= F \poisson{G}{H_2}  & 
\dot{\w} &= \phantom{\frac1{\Lambda^3} +{}} F \poisson{\w}{H_2} \\
\nonumber
\dot{K} &=0 & 
\dot{\W} &= \phantom{\frac1{\Lambda^3} +{}} F \poisson{\W}{H_2}.
\end{align}
The constants of motion are $H$, $K$ and 
\begin{equation}
\label{sd3}
C_z = -e\sin i\,\sin\w - \frac12 Fr^2 \bigbrak{1-\sin^2i \sin^2(\w+v)}. 
\end{equation}
In order to understand the geometry of the orbits for small $F$, we may
analyse the averaged Hamiltonian
\begin{equation}
\label{sd4}
\avrg{H}_\Lambda = \frac1{2\Lambda^2} - \frac32 F a e \sin i\,\sin \w.
\end{equation}
The relevant equations of motion of this one-degree-of-freedom system are 
\begin{equation}
\label{sd5}
\begin{split}
\dot{G} &= \tfrac32 F \Lambda^2 e \sin i \, \cos \w, \\
\dot{\w} &= \tfrac32 F \frac{G^4-K^2\Lambda^2}{G^3 e \sin i} \sin\w.
\end{split}
\end{equation}
We observe the existence of a pair of elliptic stationary points at $\w =
\frac\pi2, \frac{3\pi}2$ and $G^2 = \abs{K}\Lambda$, which implies $e =
\sin i = \sqrt{1-\abs{K}/\Lambda}$ (\figref{fig_s1}a). When $F=0$, the
constant of motion $C_z$ reaches its extremal values
$\pm(1-K/\abs{\Lambda})$ on these points. We will call S-orbits the
associated periodic orbits of the Stark Hamiltonian. In order to analyse
the motion at the boundaries of phase space, we use again the variables
\eqref{za7}. Since the averaged Hamiltonian can be written as
\begin{equation}
\label{sd5b}
\avrg{H_2}_\Lambda = -\frac32 \frac{\Lambda}G\x_2,
\end{equation}
they evolve according to
\begin{equation}
\label{sd6}
\begin{split}
\dot{\x_1} &= \tfrac32 F \frac{\Lambda}{G^2} \bigbrak{-\x_2^2 -
(\Lambda^2+K^2+2\x_3)\x_3} \\ 
\dot{\x_2} &= \tfrac32 F \frac{\Lambda}{G^2} \x_1\x_2 \\ 
\dot{\x_3} &= 3 F \Lambda \x_1.
\end{split}
\end{equation}
This shows that there are no other equilibrium points than the two elliptic
ones, since the flow is nonsingular at the poles (\figref{fig_s1}b).

\begin{figure}
 \centerline{\psfig{figure=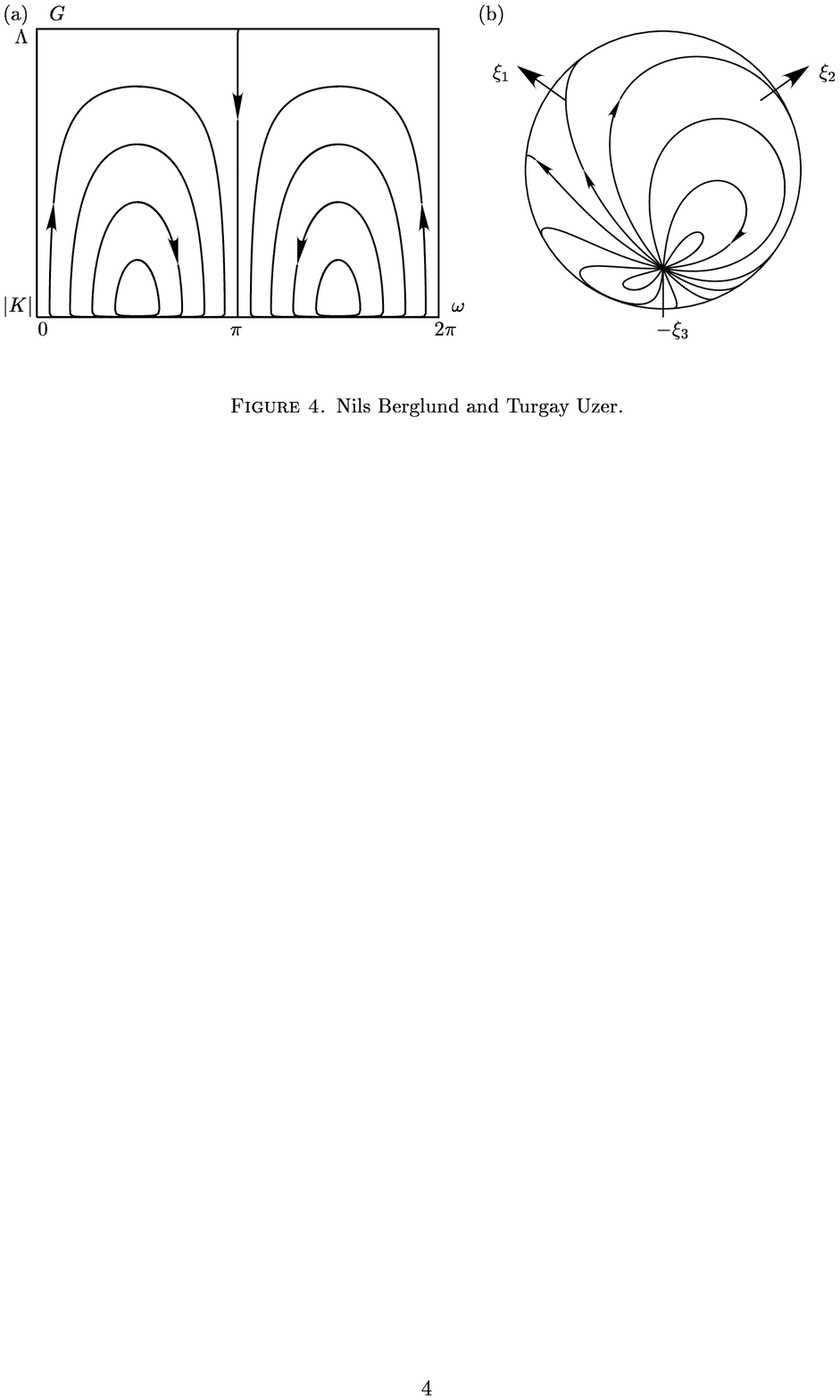,height=55mm,clip=t}}
 \vspace{1mm}
 \caption[]
 {Same as \figref{fig_s1}, but for $K=0$.}
\label{fig_s1b}
\end{figure}

Besides the periodic orbits corresponding to the elliptic equilibrium
points of \eqref{sd5}, the system displays quasiperiodic orbits for which
both $\w$ and $G$ oscillate. There is a particular orbit following the
meridian $\x_2=0$ of the sphere, for which the Kepler ellipse oscillates
between a circular and an equatorial one, while its major axis is always
perpendicular to $\vec F$. Note that the case $K=0$ is special, since the
elliptic points merge at the south pole. All orbits then go through the
south pole, where the eccentricity vanishes, which means that the electron
approaches the nucleus arbitrarily closely (\figref{fig_s1b}).

%%%%%%%%%%%%%%%%%%%%%%%%%%%%%%%%%%%%%%%%%%%%%%%%%%%%%%%%%%%%%%%%%%%%%%%%%%%%%%%%%

\subsection{Parabolic and electric action--angle variables}
\label{ssec_saa}

The separability of the Stark Hamiltonian in parabolic variables was already
known to Max Born \cite{Born} from the earlier works of P.S.\ Epstein
\cite{Epstein} and K.\ Schwarzschild \cite{Schwarzschild}. Parabolic variables
$(P_\x,P_\y,P_\ph;\x,\y,\ph)$ are defined by 
\begin{align}
\nonumber
x &= \x\y\cos\ph & 
p_x &= \frac{\y P_\x+\x P_\y}{\x^2+\y^2} \cos\ph - \frac{P_\ph}{\x\y}
\sin\ph \\
\label{saa1}
y &= \x\y\sin\ph & 
p_y &= \frac{\y P_\x+\x P_\y}{\x^2+\y^2} \sin\ph + \frac{P_\ph}{\x\y}
\cos\ph \\
\nonumber
z &= \frac12(\x^2-\y^2) & 
p_z &= \frac{\x P_\x-\y P_\y}{\x^2+\y^2},
\end{align}
where $P_\ph = L_z = K$ is the $z$-component of the angular momentum. 

\begin{figure}
 \centerline{\psfig{figure=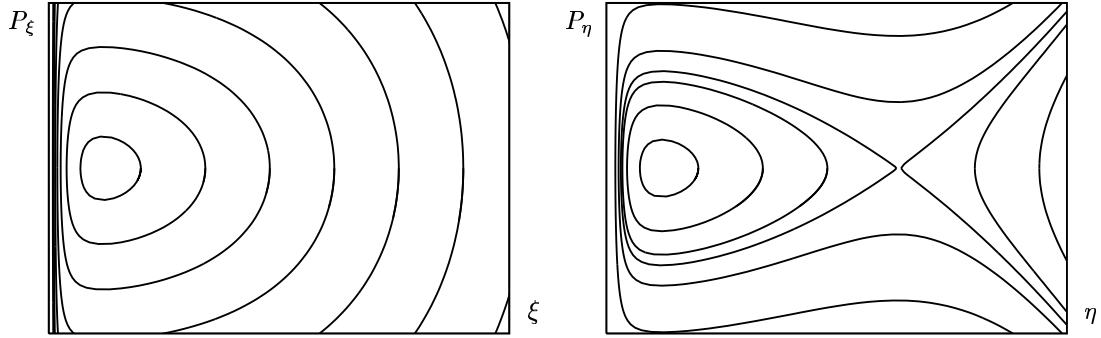,height=45mm,clip=t}}
 \vspace{1mm}
 \caption[]
 {Level lines of the constants of motion $\alpha_1$ and $\alpha_2$ for
 $H=1$, $K=\frac12$ and $F=0.05$. In contrast to $\x$, $\y$ may have an
 unbounded motion. However, the saddle point is located quite far away from
 the origin, at $\y^2=\Order{F^{-1}}$.}
\label{fig_s2}
\end{figure}

In these variables, the Hamiltonian takes the form
\begin{equation}
\label{saa2}
H = \frac1{\x^2+\y^2} \Bigbrak{\frac12 P_\x^2 + \frac{K^2}{2\x^2} + F\x^4 +
\frac12 P_\y^2 + \frac{K^2}{2\y^2} - F\y^4 - 2}.
\end{equation}
There are four constants of motion $H, \alpha_1, \alpha_2, K$, related by  
\begin{equation}
\label{saa3}
\begin{split}
\alpha_1 &= \frac12 P_\x^2 + \frac{K^2}{2\x^2} - \x^2 H + F \x^4 \\
\alpha_2 &= \frac12 P_\y^2 + \frac{K^2}{2\y^2} - \y^2 H - F \y^4 \\
\alpha_1 + \alpha_2 &= 2.
\end{split}
\end{equation}
If one scales time by a factor $\x^2+\y^2$, \eqref{saa2} is seen to 
describe the motion of two decoupled ``oscillators''. In fact, the constant
$\alpha_1$ is of the form $\frac12 P_\x^2 + V_1(\x)$, where
$V_1(\x)\to\infty$ in both limits $\x\to 0$ and $\x\to\infty$ when $F>0$.
Thus the motion of $\x$ is always bounded. By contrast, $\alpha_2 =\frac12
P_\y^2 + V_2(\y)$, where $V_2(\y)\to-\infty$ for $\y\to\infty$. Thus the
level sets of $\alpha_2$ may be unbounded for large values of $\y^2$ or
$\alpha_2$ (of order $F^{-1}$), corresponding to ionization
(\figref{fig_s2}). The following discussion is limited to the bounded
motion of $\y$, which exists for small $F$. In that case, action--angle
variables can be constructed. 

%\subsubsection{Construction of action variables}

Action variables related to the constants $\alpha_1$ and $\alpha_2$ are
defined by
\begin{equation}
\label{saa4}
\begin{split}
J_\x &= \frac1{2\pi} \oint \sqrt{2H + \frac{2\alpha_1}{\x^2} -
\frac{K^2}{\x^4} - 2F\x^2}\;\x\dx\x \\
J_\y &= \frac1{2\pi} \oint \sqrt{2H + \frac{2\alpha_2}{\y^2} -
\frac{K^2}{\y^4} + 2F\y^2}\;\y\dx\y,
\end{split}
\end{equation}
where the integrals are over bounded level sets of $\alpha_1$ and
$\alpha_2$, respectively. 
In the case $F=0$, they can be computed explicitly \cite{Born}:
\begin{equation}
\label{saa5}
J_\x = \frac12 \Bigbrak{-K + \frac{\alpha_1}{\sqrt{-2H}}}, 
\qquad
J_\y = \frac12 \Bigbrak{-K + \frac{\alpha_2}{\sqrt{-2H}}}.
\end{equation}
From this we deduce that 
\begin{equation}
\label{saa6}
H = -\frac1{2(J_\x+J_\y+K)^2},
\qquad
\alpha_{1,2} = \frac{2 J_{\x,\y} + K}{J_\x+J_\y+K},
\end{equation}
which shows, by comparison with \eqref{zd11}, that $J_\x+J_\y+K = \Lambda$
is nothing but the first Delaunay action. Moreover, one finds that the
constant of motion $C_z$ reduces to 
\begin{equation}
\label{saa7}
A_z = 1-\alpha_2 = \alpha_1-1 = \frac{J_\x-J_\y}\Lambda,
\end{equation}
which suggests to introduce the action $\Je = J_\y - J_\x$. In quantum
mechanics, $\Je$ corresponds to the electric quantum number. For further
reference, let us call $(\Lambda,\Je,K)$ the \defwd{electric action
variables}. As we have seen that $\abs{A_z}\leqs 1-\abs{K}/\Lambda$ for
$F=0$, they vary on the square domain 
\begin{equation}
\label{saa7b}
-(\Lambda-\abs{K}) \leqs \Je \leqs \Lambda-\abs{K}.
\end{equation}

For $F>0$, we need to know the expression of the Hamiltonian in action
variables. This can be done perturbatively \cite{DK83} with the result
\begin{equation}
\label{saa8}
H(\Lambda,\Je,K) = -\frac1{2\Lambda^2} - 3F\Lambda \Je -
\frac14 F^2 \Lambda^4 (17\Lambda^2 - 3 \Je^2 - 9 K^2) + \Order{F^3}. 
\end{equation}
The associated canonical equations are
\begin{align}
\nonumber
\dot{\Lambda} &= 0 & 
\dot{w}_\Lambda &= \frac1{\Lambda^3} - 3F\Je - \tfrac32 F^2 \Lambda^3
(17\Lambda^2 - 2\Je^2 - 6K^2) + \Order{F^3} \\
\label{saa9}
\dot{\Je} &= 0 & 
\dot{\we} &= \phantom{\frac1{\Lambda^3}} {}- 3F\Lambda + \tfrac32 F^2
\Lambda^4 \Je + \Order{F^3} \\ 
\nonumber
\dot{K} &= 0 &
\dot{w}_K &= \phantom{\frac1{\Lambda^3} - 3F\Lambda +{}} \tfrac92 F^2 \Lambda^4
K + \Order{F^3},
\end{align}
where $\wL$, $\we$ and $\wK$ (the \defwd{electric angle variables}) are
conjugated to $\Lambda$, $\Je$ and $K$ respectively. In the case $F=0$,
these equations are equivalent to the equations \eqref{zd12} in Delaunay
variables. The electric field suppresses the degeneracies of the Kepler
problem, and introduces different time scales for the various angles.

%\subsubsection{Construction of angle variables}

In order to compute the electric angle variables, we need to parametrize
the level curves of $\alpha_1$ and $\alpha_2$. For $F=0$, using
\eqref{saa6} the first equation of \eqref{saa3} can be written as 
\begin{equation}
\label{saa10}
P_\x^2 + \frac{(\x^2-\x_+^2)(\x^2-\x_-^2)}{\Lambda^2\x^2} = 0,
\end{equation}
where $\x_{\pm}^2 = a_1\pm b_1$ are extremal values of $\x^2$ with 
\begin{equation}
\label{saa11}
a_1 = \Lambda(2J_\x+K), 
\qquad
b_1 = 2\Lambda\sqrt{J_\x(J_\x+K)}.
\end{equation}
Similarly, $\y^2$ can vary between $a_2-b_2$ and $a_2+b_2$. In terms of the
electric actions $(\Lambda,\Je,K)$, these limits are given by
\begin{align}
\nonumber
a_1 &= \Lambda(\Lambda-\Je) & 
b_1 &= \Lambda\sqrt{(\Lambda-\Je)^2-K^2} \\
\label{saa12}
a_2 &= \Lambda(\Lambda+\Je) & 
b_2 &= \Lambda\sqrt{(\Lambda+\Je)^2-K^2}.
\end{align}
Relation \eqref{saa10} is the equation of an ellipse in the
$(\x^2,\Lambda\x P_\x)$-plane, which suggests to parametrize the level sets
by
\begin{align}
\nonumber
\x &= \sqrt{a_1 - b_1\cos\psi} & 
\y &= \sqrt{a_2 - b_2\cos\chi} \\
\label{saa13}
P_\x &= \frac{b_1\sin\psi}{\Lambda\x} & 
P_\y &= \frac{b_2\sin\chi}{\Lambda\y},
\end{align}
where $\psi$ and $\chi$ are auxiliary angles, playing a similar role as the
eccentric anomaly $E$ in the case of Delaunay variables. If we introduce the
action
\begin{equation}
\label{saa14}
S = \int^\x P_\x \dx \x' + \int^\y P_\y \dx \y' + K\ph,
\end{equation}
the angles conjugated to $(J_\x,J_\y,K)$ are given by the formulas
\begin{equation}
\label{saa15}
w_\x = \dpar{S}{J_\x}, \qquad
w_\y = \dpar{S}{J_\y}, \qquad
w_\ph = \dpar{S}{K},
\end{equation}
and the angles conjugated to $(\Lambda,\Je,K)$ are then obtained by the
linear transformation
\begin{equation}
\label{saa16}
\wL = \tfrac12(w_\x+w_\y), \qquad
\we = \tfrac12(w_\y-w_\x), \qquad
\wK = w_\ph - \wL.
\end{equation}
With the parametrization \eqref{saa13}, the derivatives \eqref{saa15} take a
simple form \cite{Born}, and the final result is
\begin{equation}
\label{saa17}
\begin{split}
\wL &= \frac{\psi+\chi}2 - \frac1{2\Lambda^2} (b_1\sin\psi + b_2\sin\chi) \\
\we &= \frac{\chi-\psi}2 \\
\wK &= \ph - \rho_1(\psi) - \rho_2(\chi),
\end{split}
\end{equation}
where 
\begin{equation}
\label{saa18}
\rho_j(\th) = \frac{K\Lambda}2 \int_0^\th \frac{\dx
\th'}{a_j-b_j\cos \th'}, \qquad j=1,2.
\end{equation}
Using the fact that $a_1^2-b_1^2 = K^2\Lambda^2$, this expression can be
written as
\begin{equation}
\label{saa19}
\rho_j(\th) = \sign(K) \Arctg
\biggbrak{\sqrt{\frac{a_j+b_j}{a_j-b_j}}\tg\frac \th2} 
\qquad \text{for $-\pi\leqs \th\leqs\pi$,}
\end{equation}
and can be continued to arbitrary $\th$ by the rule $\rho_j(\th+k2\pi) =
\rho_j(\th) + k\pi$ for $k\in\Z$. This implies that for {\em all} $\th$, we
have
\begin{equation}
\label{saa20}
\cos\rho_j(\th) = \frac{\sqrt{a_j-b_j}}{\sqrt{a_j-b_j\cos \th}} \cos\frac
\th2, 
\qquad
\sin\rho_j(\th) = \sign K \frac{\sqrt{a_j+b_j}}{\sqrt{a_j-b_j\cos \th}}
\sin\frac \th2. 
\end{equation}
Note in particular that the transformation $\we\mapsto\we+\pi$,
$\wL\mapsto\wL+\pi$ corresponds to keeping $\psi$ fixed and increasing
$\chi$ by $2\pi$. Hence it leaves all parabolic variables fixed, except
$\ph$ which is increased by $\pi$. In other words, this transformation
describes a rotation of angle $\pi$ around the $z$-axis.

The relations \eqref{saa17} are valid for $F=0$. 
Higher order expressions in $F$ of the angle variables can be computed
perturbatively, see Appendix \ref{app_aF}.

%%%%%%%%%%%%%%%%%%%%%%%%%%%%%%%%%%%%%%%%%%%%%%%%%%%%%%%%%%%%%%%%%%%%%%%%%%%%%%%%%

\subsection{Correspondence between electric action--angle and\\ 
Delaunay variables}
\label{ssec_sc}

We now establish transformation formulas between electric action--angle
variables and Delaunay variables in the case $F=0$. In the two-dimensional
case, corresponding to $K=0$, this issue has been addressed in \cite{DW91},
in connection with Lissajous variables.

The most important relations can be obtained by averaging over the fast
variable $\wL$. In view of \eqref{saa17}, the averaging operation can be
written as 
\begin{equation}
\label{sc1}
\avrg{f(\psi,\chi)}_\Lambda = 
\frac1{2\pi}\int_0^{2\pi} f(\psi,\psi+2\we) 
\Bigbrak{1 - \frac{b_1\cos\psi + b_2\cos(\psi+2\we)}{2\Lambda^2}} \dx\psi.
\end{equation}
A few useful averages are given in \tabref{t_s1}.
We can thus easily compute the averages
\begin{equation}
\label{sc2}
\begin{split}
\avrg{z}_\Lambda &= \frac12 \avrg{\x^2-\y^2}_\Lambda 
= -\tfrac32 \Lambda \Je \\
\avrg{r}_\Lambda &= \frac12 \avrg{\x^2+\y^2}_\Lambda
= \Lambda^2 \Bigbrak{1 + \frac1{8\Lambda^4} (b_1^2+b_2^2+2b_1b_2\cos2\we)}.
\end{split}
\end{equation}
Comparison with the corresponding averages over $M$ (\tabref{t_z2}) gives
us the relations
\begin{align}
\label{sc3}
\Je &= \Lambda e \sin i \sin\w,\\
\label{sc4}
e^2 &= \frac1{4\Lambda^4} \bigpar{b_1^2 + b_2^2 + 2b_1b_2\cos 2\we}.
\end{align}
This last relation allows to compute $G$ and $i$. It has a nice geometric
interpretation: the maximal value of $e^2$ (hence the minimal value of $G$)
is attained for $\we=0$ and $\pi$, while the minimal value of $e^2$ is
reached for $\we=\frac\pi2$ and $\frac{3\pi}2$. According to
\figref{fig_s1}a, these values of $\we$ correspond to
$\w=\frac\pi2\sign\Je$. From the fact that $\dot\we<0$ for $F>0$, we
infer that 
\begin{equation}
\label{sc5}
\sign(\cos\w) = -\sign(\sin2\we). 
\end{equation}
The relations \eqref{sc3}, \eqref{sc4} and \eqref{sc5} determine the
transformation $(G,\w)\mapsto(\Je,\we)$ up to a phase $\pi$ of $\we$, which
depends on $\W$. 

\begin{table}
\begin{center}
\begin{tabular}{|c|c||c|c|}
\hline
\vrule height 12pt depth 8pt width 0pt
$f$ & $\avrg{f}_\Lambda$ & $f$ & $\avrg{f}_\Lambda$ \\
\hline
\vrule height 14pt depth 8pt width 0pt
$\cos\psi$ & $-\frac{b_1}{4\Lambda^2} - \frac{b_2}{4\Lambda^2}\cos 2\we$ & 
$\cos\chi$ & $-\frac{b_1}{4\Lambda^2}\cos 2\we - \frac{b_2}{4\Lambda^2}$ \\
\vrule height 12pt depth 8pt width 0pt
$\sin\psi$ & $\frac{b_2}{4\Lambda^2}\sin 2\we$ & 
$\sin\chi$ & $-\frac{b_1}{4\Lambda^2}\sin 2\we$ \\
\vrule height 12pt depth 8pt width 0pt
$\cos^2\psi$ & $\frac12$ & 
$\cos^2\chi$ & $\frac12$ \\
\vrule height 12pt depth 8pt width 0pt
$\cos\psi\cos\chi$ & $\frac12\cos 2\we$ & 
$\cos\psi\sin\chi$ & $-\frac12\sin 2\we$ \\
\vrule height 12pt depth 9pt width 0pt
$\sin\psi\sin\chi$ & $\frac12\cos 2\we$ & 
$\sin\psi\cos\chi$ & $\frac12\sin 2\we$ \\
\hline
\end{tabular}
\end{center}
\caption[]
{The averages over the fast variable $\wL$ of some important quantities.
Using \eqref{saa13} they can be used to compute averages of some polynomials
in $\x^2$ and $\y^2$.}
\label{t_s1}
\end{table}

It remains to establish relations between the angles $(\wL,\wK)$ and
$(M,\W)$. Since $\dot{\wL} = \dot{M}$ for $F=0$, the difference $\wL-M$ does
not depend on $M$. From \eqref{saa13} we deduce that in electric
action--angle variables,
\begin{equation}
\label{sc6}
r = \Lambda^2\Bigbrak{1-\frac1{2\Lambda^2} (b_1\cos\psi+b_2\cos\chi)} =
\Lambda^2 \Bigpar{\dpar{\wL}{\psi}+\dpar{\wL}{\chi}}.
\end{equation}
Comparison with \eqref{zd6} yields the equalities
\begin{equation}
\label{sc7}
e\cos E = \frac1{2\Lambda^2} (b_1\cos\psi+b_2\cos\chi),
\end{equation}
and, using $\dot{\wL} = \dot{M}$, 
\begin{equation}
\label{sc8}
\dtot{}{E} = \dtot ME \dtot{}{M} = 
\Bigpar{\dpar{\wL}{\psi}+\dpar{\wL}{\chi}} \dtot{}{\wL} =
\dpar{}{\psi}+\dpar{}{\chi}. 
\end{equation}
Applied to \eqref{sc7}, this also gives
\begin{equation}
\label{sc9}
e\sin E = \frac1{2\Lambda^2} (b_1\sin\psi+b_2\sin\chi).
\end{equation}
In particular, when $M=0$ we have $E=0$ and thus by \eqref{saa17}
$\wL=\frac{\psi+\chi}2$. Inserting the relations $\psi=\wL-\we$ and
$\chi=\wL+\we$ into \eqref{sc7} and \eqref{sc9}, we can solve for $\cos\wL$
and $\sin\wL$ with the result, for general values of $M$,  
\begin{equation}
\label{sc10}
\cos(\wL-M) = \frac{2\Lambda^2e}{b_1+b_2} \cos\we, 
\qquad
\sin(\wL-M) = \frac{2\Lambda^2e}{b_2-b_1} \sin\we.
\end{equation}
Relation \eqref{sc8} also implies that
\begin{equation}
\label{sc11}
\psi-E = \wL-\we-M, \qquad \chi-E = \wL+\we-M.
\end{equation}

Determining $\wK$ is a bit more delicate. We will use the fact that in
Delaunay variables, the $x$-component of the angular momentum is 
\begin{equation}
\label{sc12}
L_x = \sqrt{G^2-K^2} \sin\W,
\end{equation}
while in parabolic variables, we have from \eqref{saa1}
\begin{equation}
\label{sc13}
L_x = \frac12 (\y P_\x - \x P_\y) \sin\ph 
- \frac12 \frac{K}{\x\y} (\x^2-\y^2) \cos\ph.
\end{equation}
Being independent of $M$, $L_x$ also has to be independent of $\wL$. We may
thus evaluate \eqref{sc13} in the case $\psi=0$, $\chi=2\we$. Then
\eqref{saa13} reduces to
\begin{equation}
\label{sc14}
\x=\sqrt{a_1-b_1}, \qquad
\y=\sqrt{a_2-b_2\cos2\we}, \qquad
P_\x=0, \qquad
P_\y=\frac{b_2\sin2\we}{\Lambda\y}, 
\end{equation} 
and \eqref{saa17} implies $\ph=\wK+\rho_2$, where
\begin{equation}
\label{sc15}
\cos\rho_2 = \frac{\sqrt{a_2-b_2}}\y \cos\we, \qquad
\sin\rho_2 = \sign(K)\frac{\sqrt{a_2+b_2}}\y \sin\we.
\end{equation}
Inserting \eqref{sc14} and \eqref{sc15} into \eqref{sc13}, we obtain 
\begin{equation}
\label{sc16}
L_x = L_1 \cos\wK + L_2 \sin\wK,
\end{equation}
where
\begin{align}
\nonumber
\frac{L_1}{\cos\we} &= 
-\frac1{2\Lambda\x\y^2} \Bigbrak{2\sign(K)\x^2\sqrt{a_2+b_2}\,b_2\sin^2\we +
K\Lambda(\x^2-\y^2)\sqrt{a_2-b_2}} \\
\nonumber
&= \frac12 K \sqrt{\frac{a_2-b_2}{a_1-b_1}} 
- \frac{\x}{2\Lambda\y^2\sqrt{a_2-b_2}} \Bigbrak{K\Lambda(a_2-b_2) +
2\sign(K)b_2\sqrt{a_2^2-b_2^2}\sin^2\we} \\
\label{sc17}
&= \frac12 K \biggbrak{\sqrt{\frac{a_2-b_2}{a_1-b_1}} -
\sqrt{\frac{a_1-b_1}{a_2-b_2}}\,},
\end{align}
where we have used the relation $\sqrt{a_2^2-b_2^2} = \abs{K}\Lambda$. 
The term $L_2$ can be evaluated in a similar way. We obtain the following,
relatively compact expression of $L_x$ in electric action--angle variables:
\begin{equation}
\label{sc18}
L_x = Y_1(\Lambda,\Je,K) \cos\we\cos\wK 
 + Y_2(\Lambda,\Je,K) \sin\we\sin\wK,
\end{equation}
where we have introduced the notations
\begin{equation}
\label{sc19}
\begin{split}
Y_1(\Lambda,\Je,K) &= 
\frac12 K \biggbrak{\sqrt{\frac{a_2-b_2}{a_1-b_1}} -
\sqrt{\frac{a_1-b_1}{a_2-b_2}}\,} \\
Y_2(\Lambda,\Je,K) &= 
-\frac12 \abs{K} \biggbrak{\sqrt{\frac{a_2+b_2}{a_1-b_1}} -
\sqrt{\frac{a_1-b_1}{a_2+b_2}}\,}.
\end{split}
\end{equation}
Comparison with \eqref{sc12} gives the desired relation between $\wK$ and
$\W$:
\begin{equation}
\label{sc20}
\begin{split}
\sqrt{G^2-K^2} \sin(\W-\wK) &= Y_1(\Lambda,\Je,K) \cos\we \\
\sqrt{G^2-K^2} \cos(\W-\wK) &= Y_2(\Lambda,\Je,K) \sin\we.
\end{split}
\end{equation}
We point out that despite the absolute value, the expressions \eqref{sc19}
are regular at $K=0$ and admit the Taylor series
\begin{equation}
\label{sc21}
\begin{split}
Y_1(\Lambda,\Je,K) &= -K\frac{\Je}{\sqrt{\Lambda^2-\Je^2}} 
+K^3\frac{\Lambda^2\Je}{2(\Lambda^2-\Je^2)^{5/2}} + \Order{K^5} \\
Y_2(\Lambda,\Je,K) &= -\sqrt{\Lambda^2-\Je^2} 
+K^2\frac{\Lambda^2}{2(\Lambda^2-\Je^2)^{3/2}} + \Order{K^4}.
\end{split}
\end{equation}

%%%%%%%%%%%%%%%%%%%%%%%%%%%%%%%%%%%%%%%%%%%%%%%%%%%%%%%%%%%%%%%%%%%%%%%%%%%%%%%%%

\section{The crossed-fields problem}
\label{sec_cr}

We now consider the full crossed-fields Hamiltonian \eqref{i1}, in the case
$0<B,F\ll 1$. First, we have to choose a system of coordinates. Both sets of
action--angle variables that we have used so far are defined with
respect to a privileged direction (the $z$-axis). Depending on the regime we
consider, it will be most convenient to choose this direction along the
electric or along the magnetic field. In the first case, the Hamiltonian
takes the form
\begin{equation}
\label{cr1}
H = \frac12 p^2 - \frac1r + Fz + \frac12 B L_x + \frac18 B^2(y^2+z^2).
\end{equation}
To account for the second case, we  also introduce coordinates
$(x',y',z')=(z,y,x)$. In Section \ref{ssec_cB}, we consider the case $B\ll
F$, which is a small perturbation of the integrable Stark effect, and thus
particularly well suited to perturbation theory. The case $F\ll B$ is
considered in Section \ref{ssec_cF}. The orbits contained in the plane
perpendicular to $\vec B$ exist for all values of the fields. We analyse
them in Section \ref{ssec_cq}. Other periodic orbits and the general
structure of phase space are discussed in Section \ref{ssec_cp}.

%%%%%%%%%%%%%%%%%%%%%%%%%%%%%%%%%%%%%%%%%%%%%%%%%%%%%%%%%%%%%%%%%%%%%%%%%%%%%%%%%

\subsection{The case $B\ll F$}
\label{ssec_cB}

When the magnetic field acts as a small perturbation of the Stark
Hamiltonian, it is best to use the electric action--angle variables
introduced in Section \ref{ssec_saa}. The Hamiltonian can be written as 
\begin{equation}
\label{cB1}
\begin{split}
H = H_0(\Lambda,\Je,K;F) &+ B H_1(\Lambda,\Je,K;\we,\wK;F) \\
&+ B^2 H_2(\Lambda,\Je,K;\wL,\we,\wK;F),
\end{split}
\end{equation}
where $H_0 = -\frac1{2\Lambda^2} - 3F\Lambda\Je + F^2 h_2(\Lambda,\Je,K;F)$
is the Stark Hamiltonian \eqref{saa8}, $H_1 = \frac12 L_x$ has been
computed in \eqref{sc18}, and $H_2 = \frac18(y^2+z^2)$.
The equations of motion thus have the structure
\begin{align}
\nonumber
\dot{\Lambda} &= \Order{B^2} 
& \dot{\wL} &= \frac1{\Lambda^3} - 3F\Je + \Order{F^2} + \Order{B} \\
\label{cB2}
\dot{\Je} &= \Order{B} 
& \dot{\we} &= \phantom{\frac1{\Lambda^3}} {}- 3F\Lambda + 
\Order{F^2} + \Order{B} \\
\nonumber
\dot{K} &= \Order{B} 
& \dot{\wK} &= \phantom{\frac1{\Lambda^3} - 3F\Lambda +{}} 
\Order{F^2} + \Order{B} 
\end{align}
We can again average over the fast variable $\wL$, using the rule
\eqref{sc1}. Since $L_x$ does not depend on $\wL$, $H_1$ is already in
averaged form. The equations of the averaged system are thus given by
\begin{equation}
\label{cB3}
\begin{split}
\dot{\Je} &= \phantom{-3 F\Lambda + F^2 \poisson{\we}{h_2} +{}} 
B\poisson{\Je}{H_1} + B^2\poisson{\Je}{\avrg{H_2}_\Lambda} \\
\dot{K} &= \phantom{-3 F\Lambda + F^2 \poisson{\we}{h_2} +{}}
B\poisson{K}{H_1} + B^2\poisson{K}{\avrg{H_2}_\Lambda} \\
&\\
\dot{\we} &= -3 F\Lambda + F^2 \poisson{\we}{h_2} + 
B\poisson{\we}{H_1} + B^2\poisson{\we}{\avrg{H_2}_\Lambda} \\
\dot{\wK} &= \phantom{-3 F\Lambda +{}} F^2 \poisson{\wK}{h_2} + 
B\poisson{\wK}{H_1} + B^2\poisson{\wK}{\avrg{H_2}_\Lambda} \\
\end{split}
\end{equation}
Since we assume that $B\ll F$, $\we$ evolves on a faster time scale than
$\wK$. We may thus further approximate the dynamics by averaging over $\we$,
that is, we define the double average
\begin{equation}
\label{cB4}
\avvrg{f}_{\Lambda,\Je}(K;\wK) = 
\frac1{4\pi^2} \int_0^{2\pi}\int_0^{2\pi} f(\Lambda,\Je,K;\wL,\we,\wK)
\dx \wL \dx \we.
\end{equation}
The doubly-averaged Hamiltonian $\avvrg{H}_{\Lambda,\Je}$ has one degree of
freedom and is thus integrable. 
Let us now compute various averages, at lowest order in $F$. In order
to compute $\avrg{H_2}_\Lambda$, we need to evaluate the average of
\begin{equation}
\label{cB5}
y^2+z^2 = \x^2\y^2\sin^2\ph + \tfrac14 (\x^4+\y^4-2\x^2\y^2).
\end{equation}
Using the expressions in \tabref{t_s1}, the second term is easily averaged.
To average the first term, we use the fact that with \eqref{saa17} and
\eqref{saa20}, $\x\y\sin\ph$ can be written as a polynomial in
$\sqrt{a_1\pm b_1}$, $\sqrt{a_2\pm b_2}$, and sines and cosines of $\wK$,
$\frac\psi2$ and $\frac\chi2$. The final result after simplification is
\begin{equation}
\label{cB6}
\begin{split}
\avrg{y^2+z^2}_\Lambda ={} &
\tfrac12 \Lambda^2 (2\Lambda^2 + 3\Je^2 - K^2) - \tfrac34 b_1b_2\cos 2\wK \\
&+ \tfrac14 b_1b_2 \cos 2\we - 
\tfrac12 \Lambda^2(\Lambda^2-\Je^2-K^2)\cos 2\we\,\cos 2\wK \\
&- \Lambda^2 \Je K \sin 2\we\,\sin 2\wK + \Order{F}.
\end{split}
\end{equation}

Computation of the doubly-averaged Hamiltonian is now easy. From
\eqref{sc18} and \eqref{cB6} we obtain
\begin{equation}
\label{cB7}
\begin{split}
\avvrg{L_x}_{\Lambda,\Je} &= \Order{F} \\
\avvrg{y^2+z^2}_{\Lambda.\Je} &= 
\tfrac12 \Lambda^2 (2\Lambda^2 + 3\Je^2 - K^2) - \tfrac34 b_1b_2\cos 2\wK +
\Order{F}.
\end{split}
\end{equation}
In fact, equation \eqref{aF5} in Appendix \ref{app_aF} shows that also at
order $F$ (and probably at all higher orders), the transformation
$\we\mapsto\we+\pi$, $\wL\mapsto\wL+\pi$ describes a rotation of angle
$\pi$ around the $z$-axis. Since this rotation changes $L_x$ into $-L_x$,
we conclude that $\avvrg{L_x}_{\Lambda,\Je}=\Order{F^2}$.
Discarding irrelevant constant terms, the doubly-averaged Hamiltonian can
be written as
\begin{equation}
\label{cB8}
\avvrg{H}_{\Lambda,\Je} = \tfrac94F^2\Lambda^4K^2 
-\tfrac1{16}B^2\bigbrak{\Lambda^2 K^2 + \tfrac32 b_1b_2\cos 2\wK} 
+\Order{F^3,BF^2,B^2F},
\end{equation}
where $b_1$ and $b_2$ are given in \eqref{saa12} and  the remainder denotes
a sum of terms of order $F^3$, $BF^2$ and $B^2F$. Besides $\Lambda$ and
$\Je$, \eqref{cB8} is a third adiabatic invariant of the crossed-fields
Hamiltonian in the case $B\ll F\ll 1$.  Up to the remainders, the equations
of motion have the form
\begin{equation}
\label{cB9}
\begin{split}
\dot{K} &= -\frac3{16} B^2 b_1b_2\sin 2\wK  \\
\dot{\wK} &= \frac92 F^2 \Lambda^4 K 
-\frac18 B^2\Lambda^2 K \Bigbrak{1-\frac32
\frac{\Lambda^2}{b_1b_2}(\Lambda^2+\Je^2-K^2)\cos 2\wK}.
\end{split}
\end{equation}

We should also examine the topology of phase space.  From \eqref{saa7b}, we
deduce that in the limit $F\to 0$, $K$ varies between
$-(\Lambda-\abs{\Je})$ and $\Lambda-\abs{\Je}$. If, say, $\Je\geqs 0$ and
$K=\Lambda-\Je$, we have $b_1=0$ and the angle $\psi$ is undefined. Since
$\rho_1(\psi)=\frac\psi2$ in this case, we obtain from \eqref{saa13} and
\eqref{saa17} that the quantities $\wL+\wK$ and $\wL+\we$ are sufficient to
determine the state of the system completely. We conclude that in the
averaged phase space, the variables $\wK$ and $\we$ are irrelevant when
$\abs{K} = \Lambda-\abs{\Je}$, and thus we have again a spherical topology.
The sphere can be parametrized by
\begin{equation}
\label{cB10}
(\kappa_1,\kappa_2,\kappa_3) 
= (\sqrt{(\Lambda-\abs{\Je})^2-K^2}\cos\wK,
\sqrt{(\Lambda-\abs{\Je})^2-K^2}\sin\wK, K).
\end{equation}

\begin{figure}
 \centerline{\psfig{figure=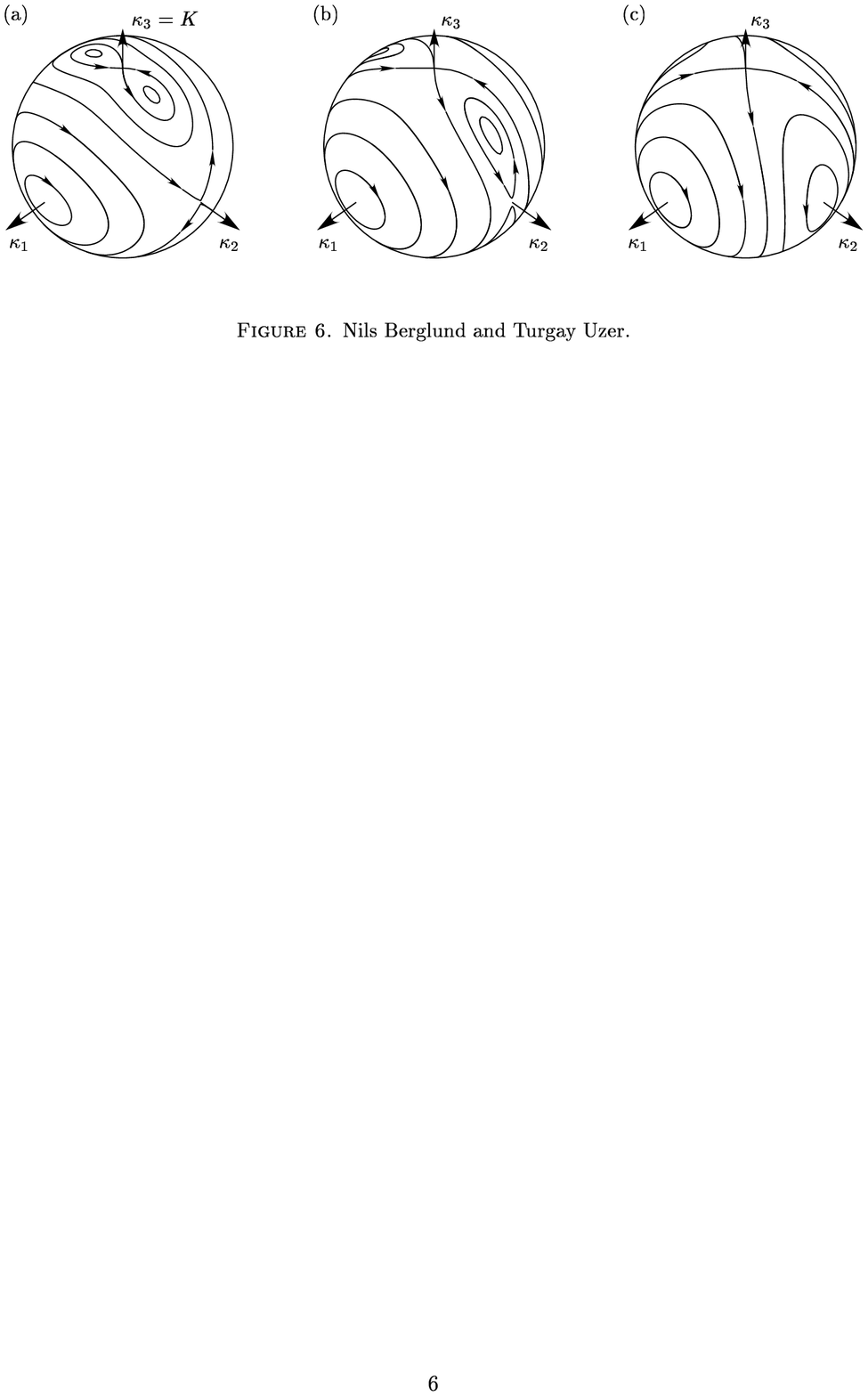,height=45mm,clip=t}}
 \vspace{1mm}
 \caption[]
 {Phase portraits of the doubly-averaged system \eqref{cB8} on the sphere
 $\kappa_1^2+\kappa_2^2+\kappa_3^2 = (\Lambda-\abs{\Je})^2$, (a) for
 $0<B<B_2$, where $B_2$ is given by \eqref{cB13}, (b) for $B_2<B<B_1$, given
 by \eqref{cB11}, and (c) for $B>B_1$. The poles (the intersection of the
 sphere with the $\kappa_3$-axis) correspond to the S-orbits present in the
 Stark effect. The intersections of the sphere with the $\kappa_1$- and
 $\kappa_2$-axis correspond respectively to the BF-orbits, lying in the
 plane defined by $\vec B$ and $\vec F$, and the B-orbits, lying in the
 plane perpendicular to $\vec B$.}
\label{fig_c1}
\end{figure}

We now analyse the structure of phase space for increasing $B$.  When
$B=0$, the orbits of \eqref{cB9} follow the parallels of the sphere.  The
poles and all points of the equator $K=0$ are fixed points.  For slightly
positive $B$, a resonance of order $2$ is created: only the points $\wK =
0,\frac\pi2, \pi, \frac{3\pi}2$ of the equator remain fixed.  A
straightforward stability analysis shows that the points $(\wK,K)=(0,0)$
and $(\pi,0)$ are always elliptic, while the points $(\frac\pi2,0)$ and
$(\frac{3\pi}2,0)$ are hyperbolic for 
\begin{equation}
\label{cB11}
B \leqs B_1 = 6\sqrt2 F\Lambda
\sqrt{\frac{\Lambda^2-\Je^2}{5\Lambda^2+\Je^2}},
\end{equation}
and elliptic for $B\geqs B_1$. If $\Je\neq 0$, there is another pair of
equilibrium points, located at $\wK=\frac\pi2, \frac{3\pi}2$ and
$K=K_\star$, given by the condition
\begin{equation}
\label{cB12}
\Bigpar{6\frac FB\Lambda}^2 = 1 + \frac32
\frac{\Lambda^2+\Je^2-K_\star^2}{\sqrt{\brak{(\Lambda-\Je)^2-K_\star^2}
\brak{(\Lambda+\Je)^2-K_\star^2}}}.
\end{equation}
These orbits are created in a pitchfork bifurcation at the poles at $B=0$
(\figref{fig_c1}a), move to the equator as $B$ increases
(\figref{fig_c1}b), and disappear in another pitchfork bifurcation at
$B=B_1$ (\figref{fig_c1}c). It is easy to see that there must be a global
bifurcation involving a saddle connection between these values. It is given
by the condition
\begin{equation}
\label{cB13}
\avvrg{H}_{\Lambda,\Je}(\tfrac\pi2,0) =
\avvrg{H}_{\Lambda,\Je}(\wK,\Lambda-\abs{\Je}) \quad\Rightarrow\quad
B = B_2 = 6\sqrt2 F\Lambda
\sqrt{\frac{\Lambda-\abs{\Je}}{5\Lambda+\abs{\Je}}}.
\end{equation}
At $B=B_2$, the poles of the sphere are connected with the points
$(\frac\pi2,0)$ and $(\frac{3\pi}2,0)$ on the equator by heteroclinic
orbits. In the case $\Je=0$, all these bifurcations collapse.

Our analysis of the doubly-averaged Hamiltonian \eqref{cB8} has thus
revealed a rather rich structure of phase space. We point out that only the
first case, depicted in \figref{fig_c1}a, is compatible with the hypothesis
$B\ll F$, which is necessary for the doubly-averaged system to be a
reliable approximation. We will see in the next sections, however, that the
picture given in \figref{fig_c1} also contains some truth in the other
parameter ranges (see \figref{fig_p3}). 

We conclude that for $B \ll F$, the structure of phase space is determined
by four types of orbits:
\begin{enum}
\item	The poles of the sphere \eqref{cB10} correspond to the fixed points
of the averaged Stark Hamiltonian, and thus to the periodic S-orbits of the
original (unaveraged) Hamiltonian \eqref{cB1}. They are unstable unless
$\Je=0$.

\item	The points $K=0$, $\wK=\frac\pi2$ or $\frac{3\pi}2$, which are
hyperbolic for small $B$, correspond to $\W=\frac\pi2$ or $\frac{3\pi}2$ in
Delaunay variables, and hence describe the B-orbits, which lie in the plane
perpendicular to $\vec B$. In the singly-averaged system, they appear as
periodic orbits, following a curve of constant $\Je$. Since $K=0$, all
these curves agglomerate at the points $G=0$ and $\w=\frac\pi2$ or
$\frac{3\pi}2$ (see \figref{fig_s1b}). In the original system, the orbits
can be interpreted as a fast rotation along a slowly ``breathing'' Kepler
ellipse, reaching periodically the eccentricity $e=1$, where the electron
approaches the nucleus indefinitely closely.

\item	The points $K=0$, $\wK=0$ or $\pi$, which are elliptic, correspond
to $\W=0$ or $\pi$, and thus describe orbits in the plane of $\vec F$ and
$\vec B$. As in the previous case, they evolve on a level curve of $\Je$
containing a point with zero angular momentum. We will call them the
BF-orbits.

\item	Finally, the four points $(\pm\frac\pi2,\pm K_\star)$ describe more
complicated orbits, which are stable and provide a connection between
S-orbits and B-orbits. For small $B$, they are close to the S-orbits. Let
us thus call them SB-orbits.
\end{enum}

%%%%%%%%%%%%%%%%%%%%%%%%%%%%%%%%%%%%%%%%%%%%%%%%%%%%%%%%%%%%%%%%%%%%%%%%%%%%%%%%%

\subsection{The case $F\ll B$}
\label{ssec_cF}

When perturbing the Zeeman effect, it seems more appropriate to use
coordinates $(x',y',z')$ in which the magnetic field is vertical. The
Hamiltonian takes the form
\begin{equation}
\label{cF1}
H = \frac12 p^2 - \frac1r + \frac12 B L_{z'} + 
\frac18 B^2 (x^{\prime2}+y^{\prime2}) 
+ F x'.
\end{equation}
We will denote by $(\Lambda',G',K';M',\w',\W')$ the associated Delaunay
variables. The transformation between these Delaunay variables and those
defined with respect to $(x,y,z) = (z',y',x')$ can be derived by expressing
$\vec L$ and $z$ in both sets of variables. The result is 
\begin{align}
\nonumber
\Lambda &= \Lambda' &
G &= G' \\
\nonumber
K &= \sqrt{G^{\prime2}-K^{\prime2}}\sin\W' &
\sin^2 i &= \cos^2\W' + \cos^2 i' \sin^2\W' \\
\label{cF2}
\cos\W &= \frac{\sqrt{G^{\prime2}-K^{\prime2}}\cos\W'}
{\sqrt{G^{\prime2}\cos^2\W'+K^{\prime2}\sin^2\W'}} & 
\cos\w &= -\frac{\sin\w'\cos\W'+\cos i'\cos\w'\sin\W'}{\sin i} \\
\nonumber
\sin\W &= \frac{K'}{\sqrt{G^{\prime2}\cos^2\W'+K^{\prime2}\sin^2\W'}} &
\sin\w &= \frac{\cos\w'\cos\W'-\cos i'\sin\w'\sin\W'}{\sin i}.
\end{align}
The averaged Hamiltonian over $M'$ takes the form
\begin{equation}
\label{cF3}
\begin{split}
\avrg{H}_\Lambda &= -\frac1{2\Lambda^2} + \frac B2 K +
B^2\avrg{H_1}_\Lambda(G',K';\w')
+ F \avrg{H_2}_\Lambda(G',K';\w',\W') \\
\avrg{H_1}_\Lambda &= \tfrac1{16} 
\bigbrak{(1+\cos^2 i')(1+\tfrac32 e^{\prime2}) +
\tfrac52 e^{\prime2} \sin^2 i' \cos 2\w'} \\
\avrg{H_2}_\Lambda &= -\tfrac32 \Lambda^2 e' (\cos\w'\cos\W' -
\sin\w'\sin\W'\cos i').
\end{split}
\end{equation}
It generates the equations of motion
\begin{align}
\nonumber
\dot{K}' &= \phantom{B^2\poisson{G'}{\avrg{H_1}_\Lambda} +{}}
F\poisson{K'}{\avrg{H_2}_\Lambda} &
\dot{\W}' &= \frac B2 + B^2\poisson{\W'}{\avrg{H_1}_\Lambda} +
F\poisson{\W'}{\avrg{H_2}_\Lambda} \\
\label{cF4}
\dot{G}' &= B^2\poisson{G'}{\avrg{H_1}_\Lambda} +
F\poisson{G'}{\avrg{H_2}_\Lambda} &
\dot{\w}' &= \phantom{\frac B2 +{}} B^2\poisson{\w'}{\avrg{H_1}_\Lambda} +
F\poisson{\w'}{\avrg{H_2}_\Lambda}. 
\end{align}
We observe that for $F\ll B$, $\W'$ evolves on a faster time scale that
$\w'$, and the dynamics can be further approximated by averaging the
Hamiltonian over $\W'$. Note, however, that the average of
$\avrg{H_2}_\Lambda$ over $\W'$ vanishes, while $\avrg{H_1}_\Lambda$ does
not depend on $\W'$. The twice averaged Hamiltonian is thus strictly
equivalent to the averaged Zeeman Hamiltonian \eqref{za4}. This means that
to lowest order in perturbation theory, the electric field does not
influence the phase portrait of the Zeeman effect, which has one of the two
behaviours indicated in \figref{fig_z2}. The phase space is thus organized
around three types of periodic orbits:
\begin{enum}
\item	a stable B-orbit at $K'=G'$, located in the plane perpendicular to
$\vec B$;
\item	a circular C-orbit, stable for large $K'$ and unstable for small $K'$;
\item	and a pair of stable Z-orbits, existing only for small $K'$, with a
major axis perpendicular to the line of nodes.
\end{enum}
These orbits will experience deformations of magnitude $\Order{F^2}$ when
$F>0$.

%%%%%%%%%%%%%%%%%%%%%%%%%%%%%%%%%%%%%%%%%%%%%%%%%%%%%%%%%%%%%%%%%%%%%%%%%%%%%%%%%

\subsection{The B-orbits}
\label{ssec_cq}

Let us now examine the most important periodic orbits for general (small)
values of $F$ and $B$. A special role is played by the orbits in the plane
perpendicular to $\vec B$, which we called B-orbits. Their existence
can already be seen on the original Hamiltonian \eqref{cr1}, which leaves
the plane $x=0$ invariant. The Hamiltonian restricted to this plane has two
degrees of freedom, its averaged version will thus have one degree of
freedom. Hence {\em all} orbits starting in that plane will look periodic in
the averaged system, but may correspond to quasiperiodic or soft chaotic
components of the original Hamiltonian.

In the two previous sections, we have found that the B-orbits are
hyperbolic in the limit $B\to 0$ and elliptic in the limit $F\to 0$. Hence
there must be at least one bifurcation value between these limits. The
twice averaged Hamiltonian in electric action--angle variables \eqref{cB8}
suggested that this transition should be given by the condition
\eqref{cB11}, which is, however, not in the range where \eqref{cB8}  can be
expected to be a good approximation. We will now examine this question in
more detail with the once averaged Hamiltonian.

\begin{figure}
 \centerline{\psfig{figure=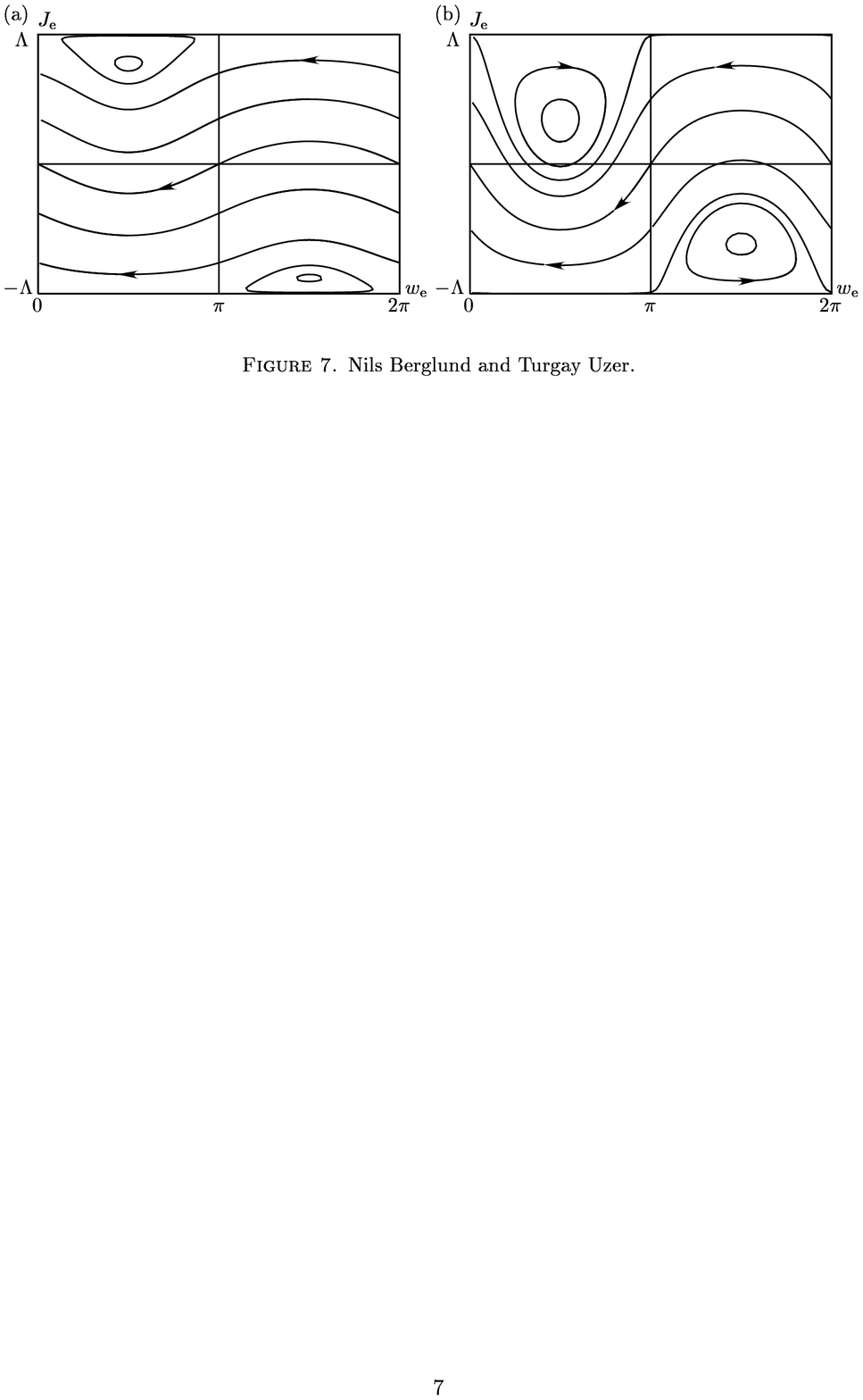,height=50mm,clip=t}}
 \vspace{1mm}
 \caption[]
 {Orbits of the averaged Hamiltonian \eqref{cq1} restricted to the plane
 perpendicular to $\vec B$. (a) shows a case with $B<F$, and (b) shows a
 case with $F<B$. There are two types of orbits, those which cross the
 lines $\we=0$ and $\pi$, and those which oscillate around $\we=\frac\pi2$
 and $\frac{3\pi}2$. The second type will not appear on a Poincar\'e
 section at $\we=0$.}
\label{fig_B1}
\end{figure}

Using the expansion \eqref{sc21} of $L_x$ for small $K$, it is easy to
compute the equations of motion \eqref{cB3} of the singly-averaged
Hamiltonian in electric action--angle variables. One can then check that
both $\dot\wK$ and $\dot K$ vanish for $K=0$ and $\cos\wK=0$, which confirms
the invariance of the subspace of B-orbits. The motion in this subspace is
determined by the one-degree-of-freedom Hamiltonian
\begin{align}
\nonumber
\avrg{H}_\Lambda(\Je,0;\we,\tfrac\pi2) 
=& -3F\Lambda\Je + \tfrac34 F^2\Lambda^4\Je^2 +\Order{F^3}\\
\nonumber
&-\tfrac12 B \sqrt{\Lambda^2-\Je^2} \sin\we + \tfrac3{32} B^2\Lambda^2
\bigbrak{\Lambda^2+\Je^2+(\Lambda^2-\Je^2)\cos 2\we} \\
&+\Order{BF}.
\label{cq1}
\end{align}
Up to the remainders, its equations of motion are given by
\begin{equation}
\label{cq2}
\begin{split}
\dot\Je &= \frac12 B\sqrt{\Lambda^2-\Je^2} \cos\we 
+ \frac3{16} B^2 \Lambda^2(\Lambda^2-\Je^2)\sin 2\we \\
\dot\we &= -3F\Lambda + \frac32 F^2\Lambda^4\Je 
+ \frac12 B \frac{\Je}{\sqrt{\Lambda^2-\Je^2}}\sin\we 
+\frac38 B^2\Lambda^2\Je\sin^2\we.
\end{split}
\end{equation}
This system admits two elliptic equilibria located at $\we=\frac\pi2$ and
$\frac{3\pi}2$ and 
\begin{equation}
\label{cq3}
\Je \simeq \pm\Lambda\Bigbrak{1+\Bigpar{\frac{B}{6F\Lambda}}^2}^{-1/2}.
\end{equation}
These points move from the boundaries $\abs{\Je}=\Lambda$ to $\Je=0$ as
$\frac BF$ increases from $0$ to $\infty$. 
\begin{figure}
 \centerline{\psfig{figure=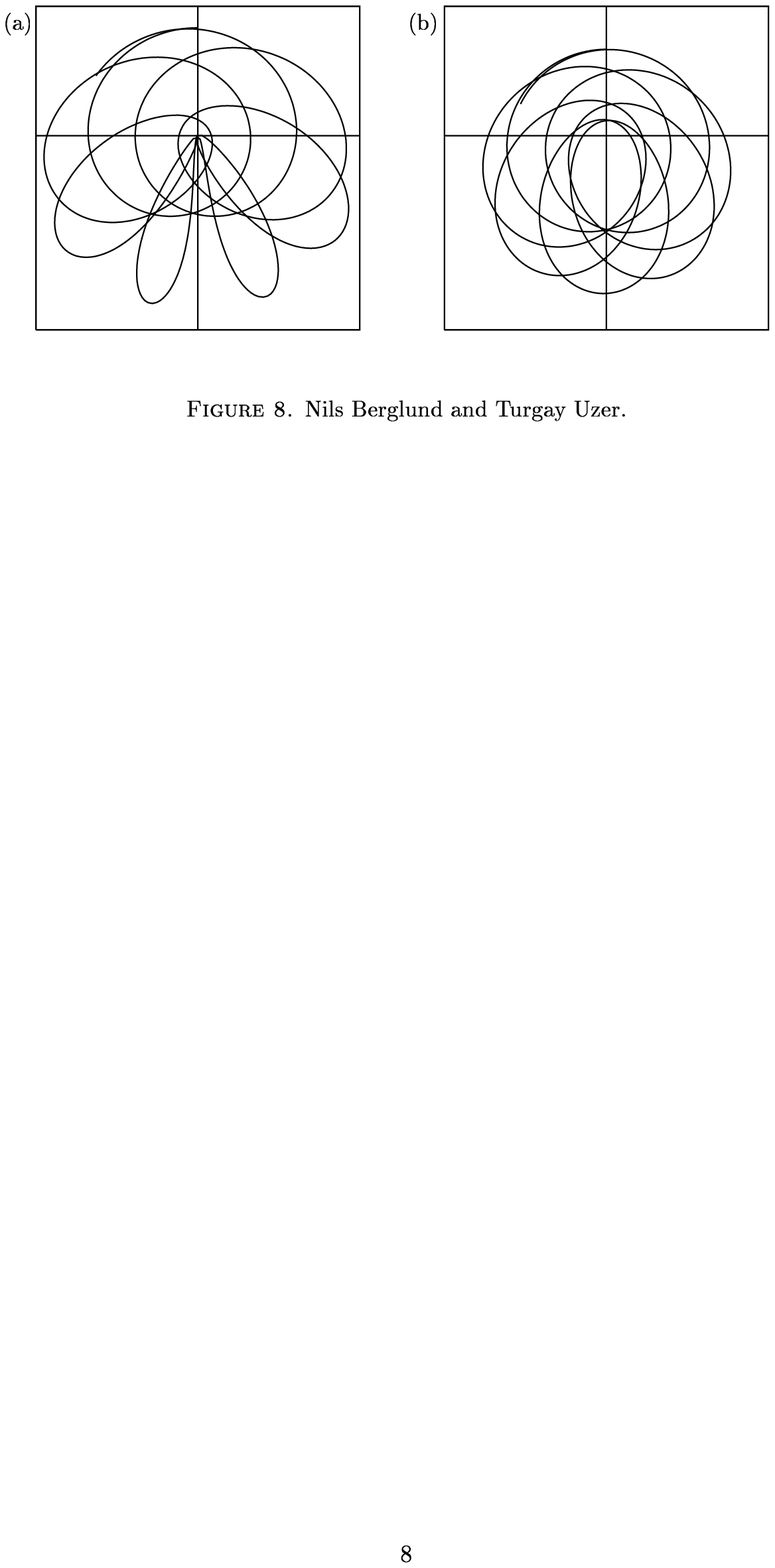,height=50mm,clip=t}}
 \vspace{1mm}
 \caption[]
 {The two types of equatorial orbits: (a) singular orbits, which
 periodically reach an eccentricity $e=1$, and (b), regular orbits which
 are bounded away from $e=1$. The electric field is vertical, while the
 magnetic field points out of the plane. One can identify the fast motion
 along a Kepler ellipse, the parameters of which evolve approximately on a
 level curve of \figref{fig_s1} in the regular case, and of
 \figref{fig_s1b} in the singular case. These orbits are not periodic in
 general, as can be seen on the pictures.}
\label{fig_B2}
\end{figure}
The orbits of \eqref{cq2} are shown in \figref{fig_B1}. They are of two
types:
\begin{enum}
\item	the orbits which cross the lines $\we=0$ and $\pi$, and for which
$\we$ is monotonous;
\item	the orbits which oscillate around $\we=\frac\pi2$ and $\frac{3\pi}2$
without reaching $\we=0$ or $\pi$. 
\end{enum}
There is no separatrix between both types of orbits, the flow being regular
when expressed in the right variables (the boundaries $\abs{\Je}=\Lambda$
should again be contracted into the poles of a sphere, since they correspond
to the periodic S-orbits of the Stark effect). Still, there is an important
qualitative difference between both types of orbits. Indeed, by \eqref{sc4},
the values $\we=0, \pi$ are the only ones that lead to an eccentricity $e=1$
when $K=0$, and thus imply a close encounter with the nucleus. Hence, the
first type of orbits will contain cusps (\defwd{singular} orbits,
\figref{fig_B2}a), and the second will not (\defwd{regular} orbits,
\figref{fig_B2}b). The boundary between both types of orbits, when starting
on the line $\we=\frac\pi2$, is given by the condition
\begin{equation}
\label{cq4}
\avrg{H}_\Lambda(\Lambda,0;\tfrac\pi2,\tfrac\pi2) 
=\avrg{H}_\Lambda(\Je,0;\tfrac\pi2,\tfrac\pi2) 
\quad\Rightarrow\quad
B\simeq 6F\Lambda\sqrt{\frac{\Lambda-\Je}{\Lambda+\Je}}.
\end{equation} 

We will now examine the stability of the $B$-orbits in the $4$-dimensional
phase space of the averaged Hamiltonian. To do this, we first need to
compute the derivatives
\begin{equation}
\label{cq5}
\begin{split}
\dpar{\dot\wK}{\wK} \Bigevalat{\frac\pi2,0} = &
\frac12 B \frac{\Je}{\sqrt{\Lambda^2-\Je^2}}\cos\we + \frac14
B^2\Lambda^2\Je\sin2\we 
= - \dpar{\dot K}{K} \Bigevalat{\frac\pi2,0}\\
\dpar{\dot\wK}{K} \Bigevalat{\frac\pi2,0} =  &
\frac92 F^2\Lambda^4 + \frac B2 \frac{\Lambda^2}{(\Lambda^2-\Je^2)^{3/2}}
\sin \we\\
&-\frac1{16} B^2 \frac{\Lambda^2}{\Lambda^2-\Je^2} 
\bigbrak{5\Lambda^2+\Je^2 + (3\Lambda^2-\Je^2)\cos 2\we} \\
\dpar{\dot K}{\wK} \Bigevalat{\frac\pi2,0} =  &
-\frac12 B\sqrt{\Lambda^2-\Je^2} \sin\we 
+ \frac18 B^2\Lambda^2(\Lambda^2-\Je^2) 
\bigbrak{3 + 2\cos 2\we}.
\end{split}
\end{equation}
In order to determine the stability of a B-orbit, we have to find the
multipliers of the variational equation
\begin{equation}
\label{cq6}
\dot{z} = A(\we,\Je)z,
\end{equation}
integrated over a solution of the system \eqref{cq2}, where $z^T =
(\wK-\frac\pi2,0)$, and $A$ is the matrix with entries given by \eqref{cq5}.

Two limiting cases can be studied analytically. When $B\to 0$, $A$ can be
replaced by its average over the fast variable $\we$, and we obtain again
the condition \eqref{cB11}, which tells us that only orbits with
$\Lambda^2-\Je^2 = \Order{B^2/F^2}$ are elliptic. The other case is that of
the orbits \eqref{cq3}, for which both $\we$ and $\Je$ are constant. In this
case, we find $\det A>0$, which means that the orbit is elliptic.

\begin{figure}
 \centerline{\psfig{figure=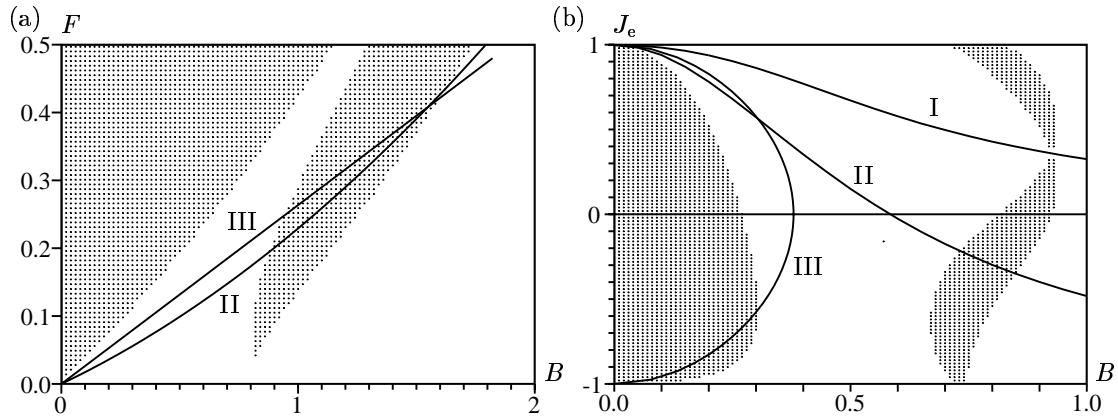,height=55mm,clip=t}}
 \vspace{1mm}
 \caption[]
 {Stability diagrams of the equatorial orbit, obtained by computing
 numerically the multipliers of equation \eqref{cq6}, starting with
 $\we=\frac\pi2$ and different values of $\Je$, $F$ and $B$. (a) shows the
 case $\Je=0$ (b) the case $F=0.1$. Shaded regions indicate unstable orbits.
 The curves are: I the location of the stable orbit \eqref{cq3}, II the
 boundary \eqref{cq4} between regular and singular orbits (orbits with large
 $B$, large $\Je$ or small $F$ are regular), and III the stability boundary
 \eqref{cB11} of the doubly-averaged system.}
\label{fig_B3}
\end{figure}

\figref{fig_B3} shows numerically computed stability diagrams for general
parameter values (one should note that equation \eqref{cq6} is much easier
to treat numerically than the full equations of motion). As expected,
orbits are unstable in a region compatible with condition \eqref{cB11} and
stable for $B\gg F$. There appears to be a second zone of instability for
intermediate values of $\frac BF$. However, this region corresponds to rather
large values of $B$, for which the averaged Hamiltonian is not necessarily a
good approximation. This confirms the picture that for small fields, the
B-orbits are stable for small $F$ and unstable for small $B$, with a linear
transition line between both regimes.

%%%%%%%%%%%%%%%%%%%%%%%%%%%%%%%%%%%%%%%%%%%%%%%%%%%%%%%%%%%%%%%%%%%%%%%%%%%%%%%%%

\subsection{The structure of phase space}
\label{ssec_cp}

The phase space of the averaged Hamiltonian is four-dimensional, and
depends on the parameters $\Lambda$, $F$ and $B$. The discussion in the
previous sections suggests that the global structure of phase space will
mainly depend on the ratio $\frac BF$; we expect that increasing both
fields, while keeping their ratio constant, will mainly result in an
increase of the size of chaotic components, without changing the location
and stability of the main periodic orbits. 

The equations of motion are given by \eqref{cB3}, where the expressions of
the linear and quadratic parts in $B$ are deduced from \eqref{sc18} and
\eqref{cB6} respectively. The manifold of constant energy is
three-dimensional, and can be represented by a Poincar\'e section. The
structure of the equations of motion shows that, at least when $\frac BF$ is not
too large, a surface of section of the form $\we=$ constant will be a good
choice.

\begin{figure}
 \centerline{\psfig{figure=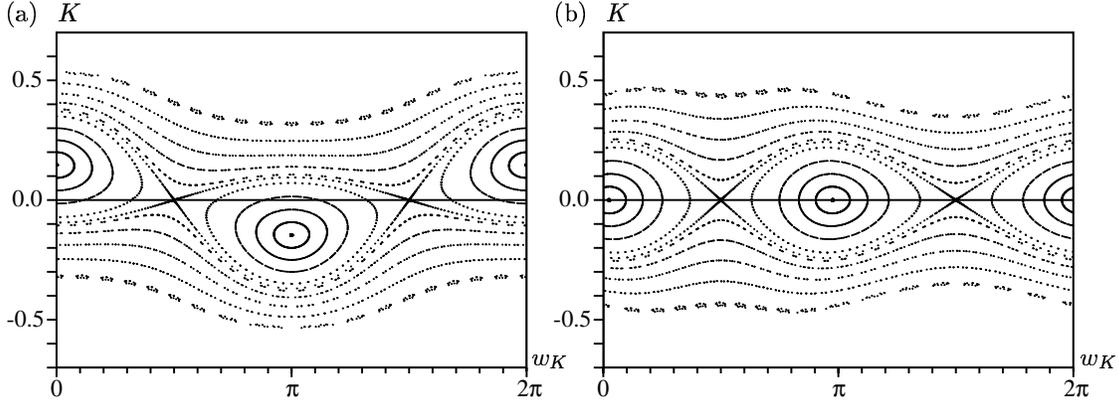,width=148mm,clip=t}}
 \vspace{1mm}
 \caption[]
 {Poincar\'e sections of the singly-averaged Hamiltonian, (a) at $\we=0$ and
 (b) at $\we=\frac\pi2$, for $B=F=0.1$, $\Lambda=1$ and $\avrg{H}_\Lambda -
 H_0 = -0.15$ (which corresponds roughly to $\Je\simeq 0.5$). Hyperbolic
 points are the B-orbits, elliptic points the BF-orbits. }
\label{fig_p1}
\end{figure}

\figref{fig_p1} shows Poincar\'e sections taken at $\we=0$ and $\frac\pi2$,
in a case with $\frac BF=1$ (sections with a smaller ratio of $\frac BF$
look similar). This section should be compared with \figref{fig_c1}a. The
hyperbolic points located at $(\wK,K)=(\frac\pi2,0)$ and $(\frac{3\pi}2,0)$
correspond to the B-orbits. Their stable and unstable manifolds separate
phase space into two regions, corresponding respectively to oscillations
around the elliptic BF-orbits and around the S-orbits. In fact, we expect
chaotic motions to show up near the separatrices, but they occupy a very
small area for these values of the fields.

In contrast to the B-orbits, the BF-orbits move in the $(\wK,K)$-plane as
$\we$ varies. This effect did not show up in the doubly-averaged
approximation, and is mainly due to terms linear in $B$ of the Hamiltonian.
\figref{fig_p1} shows that the BF-orbits are roughly located at
$K\simeq\alpha\cos\we$, $\wK\simeq\beta\sin\we$ and $K\simeq-\alpha\cos\we$,
$\wK\simeq\pi-\beta\sin\we$, where $\alpha$ and $\beta$ are of order $\frac BF$
(they can be estimated by inserting Fourier series in the equations of
motion). 

\begin{figure}
 \centerline{\psfig{figure=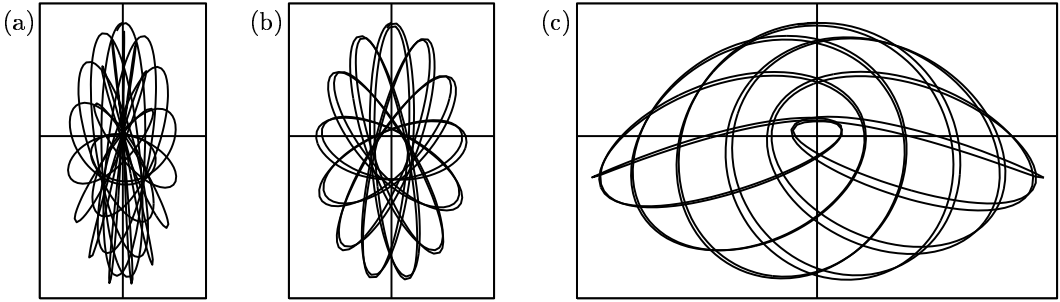,height=40mm,clip=t}}
 \vspace{1mm}
 \caption[]
 {BF-orbits, obtained by integrating the equations of motion of the
 unaveraged Hamiltonian, (a) for $F=B=0.05$ and (b,c) for $F=0.05$ and
 $B=0.1$. Initial conditions are given by \eqref{cp1} with $e=0.2$. In (a)
 and (b), $\vec F$ is vertical and $\vec B$ points out of the plane, (c)
 shows a projection on the $(\vec B,\vec F)$-plane.}
\label{fig_p2}
\end{figure}

In order to understand the geometry of BF-orbits, we observe that
\begin{enum}
\item	if $\we=0$, $K$ reaches its maximum and $\wK=0$; we know by
\eqref{sc4} that in this case the eccentricity is maximal and $\cos\w=0$;
since $K\neq 0$, however, $e$ is strictly smaller than 1, and thus the
orbit no longer approaches the nucleus, as it does for $B=0$; \eqref{sc20}
shows that $\W=\frac\pi2$ or $\frac{3\pi}2$ (depending on the sign of
$\Je$); this means that the plane of the Kepler ellipse contains $\vec
B\wedge\vec F$, and since $\cos\w=0$, its major axis is in the plane $(\vec
B,\vec F)$. 

\item	if $\we=\frac\pi2$, $K=0$ and $\wK=\beta$; we also have
$\cos\w=0$, but this time the eccentricity is minimal; \eqref{sc20}
shows that $\sin\W=\Order{\beta}$, meaning that the plane of the ellipse
contains $\vec F$, but is slightly rotated with respect to the $(\vec B,\vec
F)$-plane, by an amount of order $\frac BF$.
\end{enum}
In fact, when $\frac BF$ is small, we can deduce from \eqref{sc20} that
$\W$ is close to $0$ or $\pi$ for most values of $\we$. It approaches a
step function of the form $\frac\pi2\brak{1-\sign(\sin\we)}$ as $\frac BF$
tends to zero, which means that the orbit approaches the $(\vec B,\vec
F)$-plane in this limit. For increasing $B$, however, the orbit gains some
thickness in the direction perpendicular to the plane, and rotates around
$\vec B$ (\figref{fig_p2}). BF-orbits are thus truly non-planar when $B>0$.

Initial conditions producing BF-orbits can be constructed in the following
way. Pick an eccentricity $e\in[0,1)$. Starting with $\we=0$, we have $K=0$
and $\W=\wK=\beta$ can be read off the Poincar\'e section (though, the orbit
being elliptic, starting with a slightly wrong $\wK$ will not have dramatic
consequences). Taking $M=0$ as initial position on the ellipse, the initial
conditions are then obtained from \eqref{zd9} to be
\begin{align}
\nonumber
x &= 0 & 
p_x &= -\cos\W \tfrac1\Lambda \sqrt{\tfrac{1+e}{1-e}} \\
\label{cp1}
y &= 0 & 
p_y &= -\sin\W \tfrac1\Lambda \sqrt{\tfrac{1+e}{1-e}} \\
\nonumber
z &= \Lambda^2 (1-e) & 
p_z &= 0. 
\end{align}
We can also start with $\we=0$, $M=0$, and read $K=\alpha$ off the Poincar\'e
section, and take as initial conditions
\begin{align}
\nonumber
x &= 0 & 
p_x &= -\tfrac1\Lambda \sqrt{\tfrac{1+e}{1-e}} \\
\label{cp2}
y &= \Lambda K \sqrt{\tfrac{1+e}{1-e}} & 
p_y &= 0 \\
\nonumber
z &= \Lambda \sqrt{\Lambda^2(1-e^2) - K^2} \sqrt{\tfrac{1+e}{1-e}} & 
p_z &= 0. 
\end{align}

The boundaries of the section are given by the condition $\abs{K} = \Lambda
- \abs{\Je}$, and depend on $\wK$ and $\we$ because $\Je$ is no longer
constant. They correspond to the location of the S-orbit. The equations of
motion become singular as $\abs{K}\to\Lambda-\abs{\Je}$, which makes
unreliable the numerical computation of orbits approaching the S-orbit.
This singularity can be tamed by a canonical transformation
$(\Je,K;\we,\wK)\mapsto(\Je,J;\phi,-\wK)$. In the case $K\to\Lambda-\Je$,
$\Je\geqs 0$, for instance, it is given by  $J=\Lambda-\Je-K$ and
$\phi=\we-\wK$, the other cases being similar. We will not elaborate on
this point here.

\begin{figure}
 \centerline{\psfig{figure=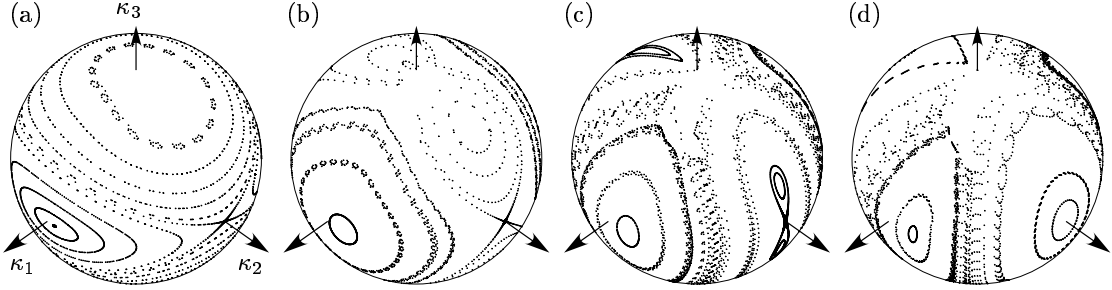,width=148mm,clip=t}}
 \vspace{1mm}
 \caption[]
 {Poincar\'e sections at $\we=\frac\pi2$ of the averaged system,
 represented on the sphere $\kappa_1^2+\kappa_2^2+\kappa_3^2 =
 (\Lambda-\abs{\Je})^2$ for increasing values of $\frac BF$, compare
 \figref{fig_c1}. In all cases, $\Lambda=1$, $\avrg{H}_\Lambda = -\frac12
 - \frac32 F$, and $B=0.1$. (a) $F=0.1$, (b) $F=0.05$, (c) $F=0.038$ and (d)
 $F=0.03$. Points are sparse near the poles, as we only plot points obtained
 with sufficient numerical accuracy.}
\label{fig_p3}
\end{figure}

\begin{figure}
 \centerline{\psfig{figure=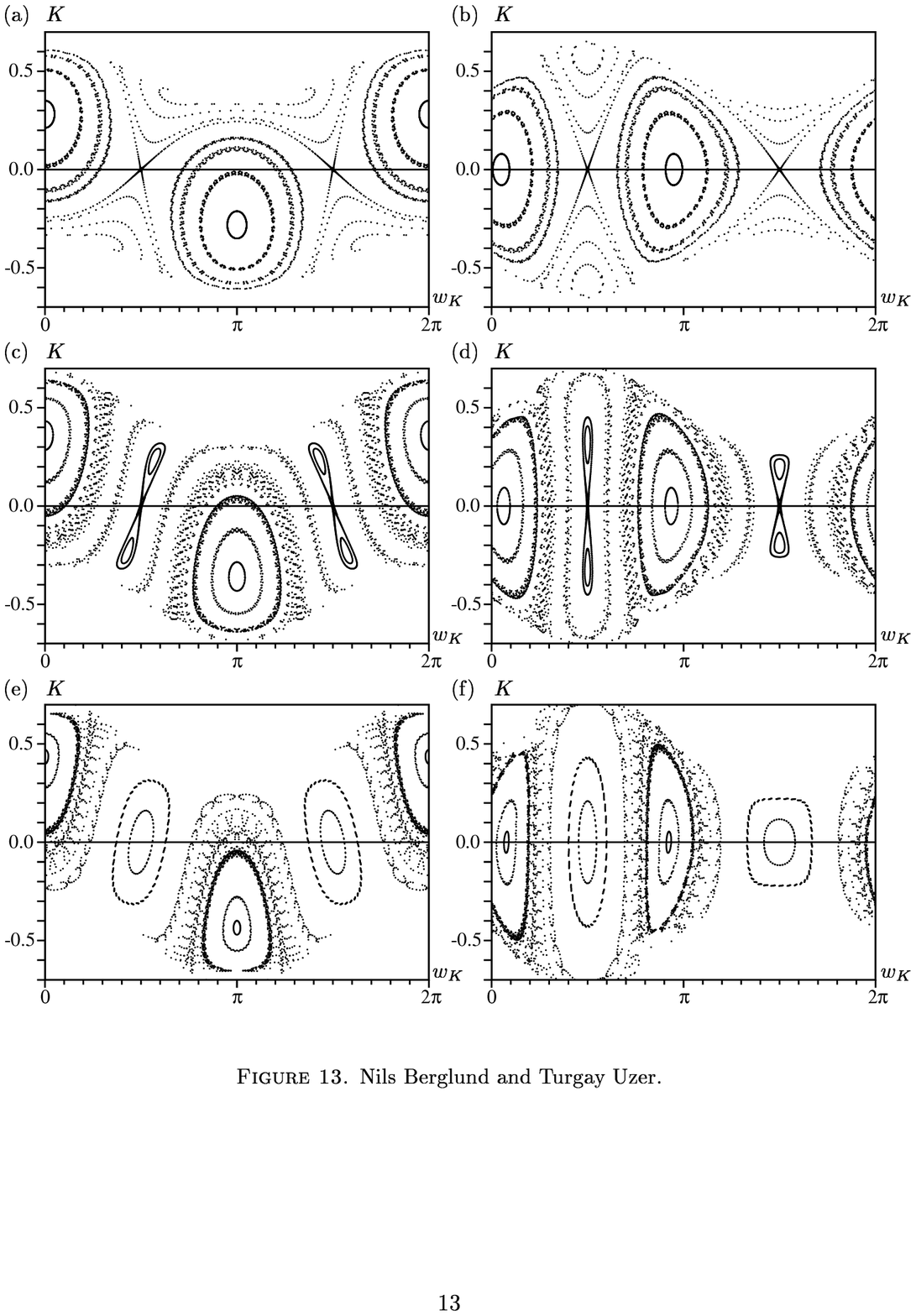,width=148mm,clip=t}}
 \vspace{1mm}
 \caption[]
 {Poincar\'e sections of the averaged Hamiltonian for increasing values of
 $\frac BF$. (a,c,e) are taken at $\we=0$ and (b,d,f) at $\we=\frac\pi2$.
 Values of $\Lambda$, $H$ and $B$ are the same as in \figref{fig_p3}, and
 (a,b) $F=0.05$, (c,d) $F=0.038$, (e,f) $F=0.03$. The equilibrium points
 $K=0$, $\wK=\frac\pi2$ and $\frac{3\pi}2$ are B-orbits, and the elliptic
 points near $\wK=0,\pi$ are BF-orbits. The SB-orbits can be recognized on
 figures (b), (c) and (d).}
\label{fig_p4}
\end{figure}

\figref{fig_p3} and \figref{fig_p4} show phase portraits for increasing
values of $\frac BF$. They mainly differ by the behaviour of the B-orbit
and its stable and unstable manifolds. While for small $\frac BF$, these
manifolds connect the B-orbits (\figref{fig_p1} and \figref{fig_p3}a), they
connect each B-orbit to an S-orbit for larger $\frac BF$ (\figref{fig_p3}b
and \figref{fig_p4}a,b). When this ratio increases further, each manifold
folds back to the B-orbit it emerges from, enclosing an elliptic orbit that
we called SB (\figref{fig_p3}c and  \figref{fig_p4}c,d). Finally, the
SB-orbits bifurcate with the B-orbit, which becomes elliptic, as we already
know (\figref{fig_p3}d and \figref{fig_p4}e,f). This scenario is very close
to the scenario obtained for the doubly-averaged system, compare
\figref{fig_c1}. 

The BF-orbit exists and remains elliptic for fairly large values of $\frac
BF$. It is possible that it exists for all values of the field, and
transforms into the Z-orbit or C-orbit of the Zeeman effect as $F\to 0$.
Verifying this conjecture would, however, require an understanding of the
transition between the Poincar\'e sections at $\we=\text{constant}$, valid
for small and moderate $\frac BF$, and those at $\W'=\text{constant}$, valid
at large $\frac BF$. At least, it is true that the geometry of BF-orbits for
$\we=0$ is compatible with that of the Z-orbits for $\W'=\frac\pi2$. 

The structure of phase space is thus essentially organized around the
B-orbits, BF-orbits and S-(or SB-)orbits, which are periodic orbits of the
averaged Hamiltonian. The other orbits of the averaged system are either
quasiperiodic with two frequencies, or belong to resonances or chaotic
components, which are small, however, for the parameter values compatible
with the averaging approximation. Intermediate values of $\frac BF$ are the
most favorable for diffusion in phase space, as the heteroclinic connections
between B- and S-orbits allow fast transitions between various regions of
phase space.

The B-orbits are called $S_+$ and $S_-$ by \cite{FW96}, who also describe a
periodic orbit located in the plane perpendicular to $\vec F$, called
$S_\perp$. We did not find such an orbit (which would correspond to $\Je=0$
and $\we=0$ or $\pi$) in the parameter range we investigated. It might be
that the BF-orbit behaves as the $S_\perp$-orbit for $F\ll B$.

Note that going from the averaged to the unaveraged dynamics will add a
time scale. Hence, periodic orbits of the averaged system may correspond to
quasiperiodic orbits of the unaveraged Hamiltonian with two frequencies, or
to soft chaotic components. Some of the quasiperiodic orbits of the
averaged system will support KAM-tori of the full system, which correspond
to a quasiperiodic motion with three frequencies.

%%%%%%%%%%%%%%%%%%%%%%%%%%%%%%%%%%%%%%%%%%%%%%%%%%%%%%%%%%%%%%%%%%%%%%%%%%%%%%%%%

\section{Conclusions and outlook}

Action--angle variables and the technique of averaging help to understand
the dynamics of the hydrogen atom in crossed electric and magnetic fields in
two ways. 

The first way is in terms of adiabatic invariants. The semi-major axis
$\Lambda^2$ of the Kepler ellipse is a constant of motion of the averaged
Hamiltonian, and thus an adiabatic invariant of the full Hamiltonian on the
time scale $B^{-2}$, compare \eqref{cr1}. The same role is played by the
averaged perturbing function $\avrg{H}_\Lambda-H_0$. In the case $B\ll F$, we
have the additional invariants $\Je$ and $\avvrg{H}_{\Lambda,\Je}$ (see
\eqref{cB6}), which evolve on the time scale $B^{-1}$. In the case $F\ll
B$, $K'=\vec L\cdot\vec B/B$ and the averaged quadratic Zeeman term are
adiabatic invariants on the time scale $F^{-1}$.

The second way is in terms of periodic orbits of the averaged system, which
organize the structure of phase space. The B-orbits, contained in the plane
perpendicular to $\vec B$, exist for all fields. They are unstable for
$B\to 0$ and stable for $F\to 0$, with a roughly linear transition line in
the $(F,B)$-plane. In the limit $B\to 0$, we also find the S-orbits of the
Stark effect, and the stable BF-orbits contained in the plane of $\vec B$
and $\vec F$. As $B$ increases, S-orbits appear to become unstable by
expelling a pair of ``SB-orbits'', which are then absorbed by the B-orbits.
BF-orbits  are non-planar for $B>0$. They  exist and are stable in a large
domain of field values. It is unclear whether they transform into one of
the periodic orbits of the pure Zeeman effect as $F\to 0$.

Diffusion in phase space is most prominent in the regime where $F$ and $B$
are of comparable magnitude, when neighbourhoods of the S-orbits are
connected by heteroclinic orbits. In the other regions of phase space,
close to the stable orbits,  orbits of the averaged system are trapped
inside KAM-tori. In the real three-degrees-of-freedom system, however,
diffusion becomes possible in these regions as well, although on a longer
time scale. This aspect of dynamics still has to be better understood, as
do the  quantum and semiclassical mechanics of the problem which continue
to be of interest \cite{vMU97,vMFU97b,NUFESWHD97,RT97,CZRT99}.

%%%%%%%%%%%%%%%%%%%%%%%%%%%%%%%%%%%%%%%%%%%%%%%%%%%%%%%%%%%%%%%%%%%%%%%%%%%%%%%%%

\section*{Acknowledgments}

This work originated during a very pleasant stay of the first author in the
group of Jacques Laskar, at the Bureau des Longitudes in Paris. We thank
him for introducing us to the realm of perturbed Kepler problems, Delaunay
variables, and averaging. NB was supported by the Fonds National Suisse de
la Recherche Scientifique.

%%%%%%%%%%%%%%%%%%%%%%%%%%%%%%%%%%%%%%%%%%%%%%%%%%%%%%%%%%%%%%%%%%%%%%%%%%%%%%%%%

\appendix

\section{Averaging}
\label{app_av}

A standard result on averaging \cite{V96} is

\begin{theorem}
%\label{}
Consider the initial value problem
\begin{equation}
\label{av1}
\dot{x} = \eps f(x,t) + \eps^2 g(x,t,\eps), 
\qquad x(0)=x_0,
\end{equation}
where $f(x,t)$ is $2\pi$-periodic in $t$, and $x, x_0\in\R^n$. Define the
averaged system
\begin{equation}
\label{av2}
\begin{split}
\dot{y} &= \eps\avrg{f}(y), 
\qquad y(0)=x_0,\\
\avrg{f}(y) &= \frac1{2\pi} \int_0^{2\pi} f(y,t)\dx t.
\end{split}
\end{equation}
If $f$ and $g$ are sufficiently smooth and bounded, then
\begin{enum}
\item	$x(t)-y(t) = \Order{\eps}$ on the time scale $1/\eps$;
\item	If $y_0$ is a nondegenerate equilibrium of \eqref{av2}, then
\eqref{av1} possesses an isolated periodic orbit
$\gamma(t)=y_0+\Order{\eps}$, with the same stability as $y_0$ if $y_0$ is
hyperbolic.
\end{enum}
\end{theorem}

Consider now a Hamiltonian of the form
\begin{equation}
\label{av3}
H(I_1,\dots,I_n;\ph_1,\dots,\ph_n) 
= H_0(I_1) + \eps H_1(I_1,\dots,I_n;\ph_1,\dots,\ph_n),
\end{equation}
where the $\ph_i$ are angle variables.
If we introduce the Poisson bracket
\begin{equation}
\label{av4}
\poisson{f}{g} = \sum_{i=1}^n \Bigbrak{\dpar f{\ph_i} \dpar g{I_i} - \dpar
f{I_i} \dpar g{\ph_i}},
\end{equation}
the equations of motion can be written in the form
\begin{align}
\nonumber
\dot{\ph_1} &= H_0'(I_1) + \eps \poisson{\ph_1}{H_1} &&\\
\label{av5}
\dot{\ph_j} &= \eps \poisson{\ph_j}{H_1} &\qquad j=&2,\dots,n\\
\nonumber
\dot{I_i} &= \eps \poisson{I_i}{H_1} &\qquad i=&1,\dots,n.
\end{align}
If $\poisson{\ph_1}{H_1}$ is bounded, $H_0'(I_1)\neq 0$ and $\eps$ is small
enough, we may reparametrize the orbits by $\ph_1$ instead of $t$, giving
\begin{equation}
\label{av6}
\begin{split}
\dtot{\ph_j}{\ph_1} &= \eps \frac{\poisson{\ph_j}{H_1}}{H_0'(I_1)} +
\Order{\eps^2} \qquad j=2,\dots,n\\
\dtot{I_i}{\ph_1} &= \eps \frac{\poisson{I_i}{H_1}}{H_0'(I_1)} +
\Order{\eps^2} \qquad i=1,\dots,n.
\end{split}
\end{equation}
According to the theorem, the dynamics of this system are well approximated
by those of the system averaged over the fast variable $\ph_1$,
\begin{equation}
\label{av7}
\begin{split}
\dtot{\ph_j}{\ph_1} &= \eps \frac{\poisson{\ph_j}{\avrg{H_1}}}{H_0'(I_1)}
\qquad j=2,\dots,n\\
\dtot{I_i}{\ph_1} &= \eps \frac{\poisson{I_i}{\avrg{H_1}}}{H_0'(I_1)}
\qquad i=1,\dots,n,
\end{split}
\end{equation}
where we have introduced
\begin{equation}
\label{av8}
\avrg{H_1}(I_1,\dots,I_n;\ph_2,\dots,\ph_n) =
\frac1{2\pi}\int_0^{2\pi} H_1(I_1,\dots,I_n;\ph_1,\dots,\ph_n) \dx \ph_1.
\end{equation}
We now observe that the canonical equations associated with the averaged
Hamiltonian $\avrg{H} = H_0+\eps\avrg{H_1}$,
\begin{align}
\nonumber
\dot{\ph_1} &= H_0'(I_1) + \eps \poisson{\ph_1}{\avrg{H_1}} &&\\
\nonumber
\dot{I_1} &= 0 && \\
\label{av9}
\dot{\ph_j} &= \eps \poisson{\ph_j}{\avrg{H_1}} & j&= 2,\dots n \\
\nonumber
\dot{I_j} &= \eps \poisson{I_j}{\avrg{H_1}} & j&= 2,\dots n
\end{align}
are close to $\Order{\eps^2}$ to those of the averaged system \eqref{av7}.
Thus the dynamics of the averaged Hamiltonian are a good approximation, in
the sense of the averaging theorem, of those of the initial Hamiltonian. In
particular, the averaged Hamiltonian \eqref{av8} is an adiabatic invariant
of the initial system.

The averaging procedure can be extended to higher orders in $\eps$ by the
method of Lie-Deprit series \cite{Deprit,Henrard}.

%%%%%%%%%%%%%%%%%%%%%%%%%%%%%%%%%%%%%%%%%%%%%%%%%%%%%%%%%%%%%%%%%%%%%%%%%%%%%%%%%

\begin{table}
\begin{center}
\begin{tabular}{|c|ccccc|}
\hline
%\poisson{\cdot}{\cdot} 
\vrule height 13pt depth 7pt width 0pt
& $\Lambda$ && $M$ && $\w$ \\
\hline
\vrule height 20pt depth 10pt width 0pt
$e$ & 
$0$ && $\dfrac{G^2}{\Lambda^3 e}$ && 
$-\dfrac{G}{\Lambda^2 e}$  \\ 
\vrule height 20pt depth 10pt width 0pt
$r$ & 
$-\dfrac{\Lambda^4 e}{r} \sin E$ && 
$\dfrac{2r}{\Lambda} - \dfrac{G^2}{\Lambda e}\cos E$ && 
$\dfrac{G}{e} \cos E$ \\
\vrule height 20pt depth 10pt width 0pt
$\sin^2 i$ & $0$ && $0$ && $2\dfrac{K^2}{G^3}$ \\
\vrule height 20pt depth 10pt width 0pt
$v$ & 
$-\dfrac{\Lambda^3 G}{r^2}$ && 
$\dfrac{G\sin E}{re}$ && 
$-\dfrac{\Lambda}{re} \sin E$ \\
\vrule height 20pt depth 10pt width 0pt
$X$ & 
$\dfrac{\Lambda^4}{r}\sin E$ &&
$-\dfrac{r^2}{e\Lambda^3} - \Lambda e \sin^2 E$ && 
$\dfrac Ge$ \\
\vrule height 20pt depth 12pt width 0pt
$Y$ & 
$-\dfrac{\Lambda^3}{r}G\cos E$ && 
$G \sin E$ && $\Lambda \sin E$ \\
\hline
\end{tabular}
\end{center}
\caption[]
{A few useful Poisson brackets involving Delaunay variables. We show
Poisson brackets between columns and lines, for instance the upper right
element is $\poisson{\w}e$. Brackets involving $G$, $K$ and $\W$ are
trivial to compute.}
\label{t_z1}
\end{table}

\section{Poisson brackets}
\label{app_pb}

In this appendix, we discuss the computation of Poisson brackets involving
various sets of variables used in this paper.

Since Hamiltonians are usually expressed in terms of auxiliary variables,
such as the eccentricity $e$ or inclination $i$, it is useful to start by
computing some Poisson brackets involving these quantities. Some of them are
shown in \tabref{t_z1}. 

For instance, for the perturbation term $H_1=\frac1{16} r^2\bigbrak{1 +
\cos^2 i + \sin^2 i \,\cos(2\w+2v)}$ of the Zeeman Hamiltonian, we find 
\begin{align}
\nonumber
\poisson{\Lambda}{H_1} =& -\frac18 \Lambda^4 e 
\bigbrak{1 + \cos^2 i + \sin^2 i \,\cos(2\w+2v)} \sin E 
+ \frac18 \Lambda^3 G \sin^2 i \,\sin(2\w+2v) \\
\nonumber
\poisson{G}{H_1} =& \frac18 r^2 \sin^2 i\,\sin(2\w+2v) \\
\nonumber
\poisson{M}{H_1} =& \frac18 r \Bigpar{\frac{2r}\Lambda - 
\frac{G^2}{\Lambda e}\cos E} 
\bigbrak{1 + \cos^2 i + \sin^2 i \,\cos(2\w+2v)} \\
\nonumber
&- \frac18 \frac{rG}e \sin^2 i \,\sin(2\w+2v) \sin E \\
\nonumber
\poisson{\w}{H_1} =& \frac18 \frac{rG}{e} 
\bigbrak{1 + \cos^2 i + \sin^2 i \,\cos(2\w+2v)} \cos E \\
&+ \frac18 \frac{r\Lambda}e \sin^2 i \,\sin(2\w+2v) \sin E 
+ \frac18 \frac{r^2}{G} \bigbrak{-1 + \cos(2\w+2v)}.
\label{pb1}
\end{align}
There is an apparent singularity at $e=0$. It can be removed, however, by
introducing variables $J=\Lambda-G=\Order{e^2}$ and $\phi=M+\w$. The
transformation $(\Lambda,G,K;M,\w,\W)\mapsto(\Lambda,J,K;\phi,-\w,\W)$ is
canonical. One can check that $\dot\phi$ is finite in the limit $e\to 0$,
and that the variable $\z=J\e^{\icx\w}$ satisfies $\dot\z=\Order{e}$. This
shows that the circular orbit is indeed a periodic orbit of the Zeeman
Hamiltonian, as suggested by the averaged system.

Another way to deal with the singularity of Delaunay variables at $e=0$ in
the averaged case is to use coordinates $(\x_1,\x_2,\x_3)$ introduced by
\cite{CDMW87}, see equation \eqref{za7}. \tabref{t_z3} gives some useful
Poisson brackets involving these variables.

\begin{table}
\begin{center}
\begin{tabular}{|c|ccccc|}
\hline 
\vrule height 14pt depth 8pt width 0pt
 & $\x_1$ && $\x_2$ && $\x_3$ \\
\hline 
\vrule height 14pt depth 8pt width 0pt
$\x_1$ & $0$ && $-2G\x_3$ && $2G\x_2$ \\
\vrule height 12pt depth 8pt width 0pt
$\x_2$ & $2G\x_3$ && $0$ && $-2G\x_1$ \\
\vrule height 12pt depth 8pt width 0pt
$\x_3$ & $-2G\x_2$ && $2G\x_1$ && $0$ \\
\hline 
\vrule height 14pt depth 8pt width 0pt
$G$ & $-\x_2$ && $\x_1$ && $0$ \\
\vrule height 14pt depth 12pt width 0pt
$e^2$ & $2\dfrac{1-e^2}G \x_2$ && $-2\dfrac{1-e^2}G \x_1$ && $0$ \\
\vrule height 14pt depth 12pt width 0pt
$\cos^2 i$ & $2\dfrac{\cos^2 i}G \x_2$ && $-2\dfrac{\cos^2 i}G \x_1$ && $0$ \\
\hline
\end{tabular}
\end{center}
\caption[]
{Some Poisson brackets involving the variables $\x_i$ defined in
\eqref{za7}.}
\label{t_z3}
\end{table}

%%%%%%%%%%%%%%%%%%%%%%%%%%%%%%%%%%%%%%%%%%%%%%%%%%%%%%%%%%%%%%%%%%%%%%%%%%%%%%%%%

\section{Angle variables to first order in $F$}
\label{app_aF}

For $F>0$, the extremal values of $\x^2$ are given by the condition
\begin{equation}
\label{aF1}
K^2 - 2\alpha_1(F)\x^2 - 2H(F)\x^4 + 2 F \x^6 = 0,
\end{equation}
where the quantities
\begin{equation}
\label{aF2}
\begin{split}
H(F) &= -\frac1{2\Lambda^2} - 3F\Lambda\Je + \Order{F^2} \\
\alpha_{1,2}(F) &= \frac{2J_{\x,\y}+K}{\Lambda} 
\pm F \Lambda^2 \bigbrak{6J_\x J_\y + 3K(J_\x+J_\y)+K^2} + \Order{F^2}
\end{split}
\end{equation}
are obtained by solving equations \eqref{saa4} perturbatively. We find that
$\x^2$ varies between limits $\hat a_1(F) \pm \hat b_1(F)$ given by
\begin{equation}
\label{aF3}
\begin{split}
\hat a_1(F) &= a_1 + \tfrac12 F \Lambda^4
\bigbrak{\Je^2+4\Lambda\Je-5\Lambda^2+K^2}  + \Order{F^2}\\
\hat b_1(F) &= b_1
\bigbrak{1-\tfrac12F\Lambda^3(5\Lambda+\Je)}  + \Order{F^2}.
\end{split}
\end{equation}
Likewise, the bounded orbits of $\y^2$ vary between limits $\hat a_2(F) \pm
\hat b_2(F)$ which are obtained by changing the signs of $\Je$ and $F$ in
\eqref{aF3}. We may parametrize the level curves of $\alpha_{1,2}$ by 
\begin{equation}
\label{aF4}
\begin{split}
\x^2 &= \hat a_1(F) - \hat b_1(F)\cos\psi \\
\y^2 &= \hat a_2(F) - \hat b_2(F)\cos\chi,
\end{split}
\end{equation}
with the momenta given by \eqref{saa3}. The derivatives of the action
\eqref{saa14} can then be computed as before, with the result
\begin{equation}
\label{aF5}
\begin{split}
\wL =& \frac{\psi+\chi}2 - \frac1{2\Lambda^2} (b_1\sin\psi+b_2\sin\chi) \\ 
&+ \tfrac14 F \bigbrak{\Lambda(13\Lambda+7\Je)b_1\sin\psi - b_1^2\sin 2\psi
- \Lambda(13\Lambda-7\Je)b_2\sin\chi + b_2^2\sin 2\chi}\\
&+\Order{F^2} \\
\we =& \frac{\chi-\psi}2 + F\Lambda^2\bigbrak{b_1\sin\psi+b_2\sin\chi} 
+\Order{F^2} \\
\wK =& \ph - \rho_1(\psi) - \rho_2(\chi) 
+ \frac12 F K\Lambda^4 \Bigbrak{\frac{\sin\psi}{a_1-b_1\cos\psi} -
\frac{\sin\chi}{a_2-b_2\cos\chi}} +\Order{F^2}.
\end{split}
\end{equation} 

%%%%%%%%%%%%%%%%%%%%%%%%%%%%%%%%%%%%%%%%%%%%%%%%%%%%%%%%%%%%%%%%%%%%%%%%%%%%%%%%%


\begin{thebibliography}{vMFU97b}

\small

%% New definition of \@listI for the bibliography
\makeatletter
\def\@listI{\leftmargin\leftmargini
        \topsep=\medskipamount
        \setlength{\parsep}{0mm}
        \setlength{\itemsep}{-1.5mm}
}
\let\@listi\@listI
\@listi
\makeatother

\bibitem[B27]{Born}	\bibbook{M.\ Born}
			{The Mechanics of the Atom}
			{G.\ Bell}
			{London, 1927}

\bibitem[BM75]{Bohr/Mottelson}  
			\bibbook{A.\ Bohr, B.R.\ Mottelson}
			{Nuclear Structure, Vol. II}
			{Benjamin Reading}{MA 1975}

\bibitem[BS84]{Braun/Solovev}  
			\bibarticle{P.A.\ Braun, E.A.\ Solov'ev}
			{The Stark Effect for a hydrogen atom in a magnetic
			field}
			{Sov.\ Phys.\ JETP}
			{59}{38}{46}{1984}
			
\bibitem[C98]{Connerade}
			\bibbook{J.-P.\ Connerade}
			{Highly Excited Atoms}
			{Cambridge U.P.}
			{Cambridge, UK, 1998}
			
\bibitem[D69]{Deprit}	\bibarticle{A.\ Deprit}
			{Canonical transformations depending on a small
			parameter}
			{Celestial Mech.}{1}{12}{30}{1969}

\bibitem[CDMW87]{CDMW87}	
			\bibarticle{S.L.\ Coffey, A.\ Deprit, B.\ Miller,
			C.A.\ Williams} 
			{The quadratic Zeeman Effect in Moderately Strong
			Magnetic Fields}
			{Ann.\ N.Y.\ Acad.\ Sci.}{497}{22}{36}{1987}

\bibitem[CZRT99]{CZRT99}
			\bibarticle{J.-P.\ Connerade, M.-S.\ Zhan, J.\ Rao,
			K.T.\ Taylor}
			{Strontium spectra in crossed electric and magnetic
			fields}
			{J.\ Phys.\ B}{32}{2351}{2360}{1999}
			
\bibitem[DG89]{Delande/Gay}	
			\bibtitle{D.\ Delande, J.-C.\ Gay} 
			{Quantum Chaos and the Hydrogen Atom in Strong
			Magnetic Fields}
			in \bibbook{G.F.\ Bassani, M.\ Inguscio, T.W.\
			H\"ansch Eds.}
			{The Hydrogen Atom}
			{Springer-Verlag}{Berlin, 1989}
			
\bibitem[DK83]{DK83}	\bibtitle{R.J.\ Damburg, V.V.\ Kolosov}
			{Theoretical studies of hydrogen Rydberg atoms in
			electric fields}
			in \bibbook{R.F.\ Stebbings, F.B.\ Dunning Eds.} 
			{Rydberg states of atoms and molecules}
			{Cambridge Univ.\ Press}
			{Cambridge, 1983}
			
\bibitem[DKN83]{DKN83}	\bibarticle{J.B.\ Delos, S.K.\ Knudson, D.W.\ Noid} 
			{Highly excited states of a hydrogen atom in a
			strong magnetic field}
			{\PRA}{28}{7}{21}{1983}
			
\bibitem[DS92]{Digman/Sipe}  
			\bibarticle{M.M.\ Dignam, J.E.\ Sipe}
			{Semiconductor superlattice exciton states in
			crossed electric and magnetic fields}
			{\PRB}{45}{6819}{6838}{1992}

\bibitem[DW91]{DW91}	\bibarticle{A.\ Deprit, C.A.\ Williams} 
			{The Lissajous transformation. IV. Delaunay and
			Lissajous variables}
			{Celestial\ Mech.\ Dynam.\ Astronom.}
			{51}{271}{280}{1991}
			
\bibitem[E16]{Epstein}	{P.S.\ Epstein},
			{}
			\bibref{Ann.\ Phys.}{50}{489}{}{1916}
			\bibref{Ann.\ Phys.}{58}{553}{}{1919}.
			
\bibitem[F94]{Farrelly} 
		 	\bibarticle{D.\ Farrelly}
			{Motional Stark effect on Rydberg states in crossed
			electric and magnetic fields}
			{Phys.\ Lett.\ A}{191}{265}{}{1994}
			
\bibitem[F\&92]{Farrelly/Uzer/&92}
			\bibarticle{D.\ Farrelly, T.\ Uzer, P.E.\ Raines,
			J.P.\ Sketton, J.A.\ Milligan}
			{Electronic structure of Rydberg atoms in parallel
			electric and magnetic fields}
			{\PRA}{45}{4738}{4751}{1992}

\bibitem[FlWe96]{FW96}	\bibarticle{E.\ Fl\"othmann, K.H.\ Welge}
			{Crossed-field hydrogen atom and the three-body
			Sun-Earth-Moon problem}
			{\PRA}{54}{1884}{1888}{1996}

\bibitem[FrWi89]{Friedrich/Wintgen}  
			\bibarticle{H.\ Friedrich, D.\ Wintgen}
			{The hydrogen atom in a uniform magnetic field - an
			example of chaos}
			{Phys.\ Rep.}
			{183}{37}{}{1989}
			
\bibitem[G90]{Gutzwiller}  
			\bibbook{M.C.\ Gutzwiller}
			{Chaos in Classical and Quantum Mechanics}
			{Springer-Verlag}
			{New York, 1990}
			
\bibitem[G98]{Gutzwiller2}
			\bibarticle{M.C.\ Gutzwiller}
			{Moon-Earth-Sun: The oldest three-body problem}
			{\RMP}{70}{589}{639}{1998}

\bibitem[GT69]{Garton/Tomkins}  
			{W.R.S.\ Garton, F.S.\ Tomkins},
			{}
			\bibref{Astrophys.\ J.}{158}
			{839}{}{1969}.
			
\bibitem[H70]{Henrard}
			\bibarticle{J.\ Henrard}
			{On a perturbation theory using Lie transforms}
			{Celestial Mech.}{3}{107}{120}{1970}

\bibitem[HRW89]{Hasegawa/Robnik/Wunner}
			\bibarticle{H.\ Hasegawa, M.\ Robnik, G.\ Wunner}
			{Classical and quantal chaos in the diamagnetic {K}epler
             		problem}
			{Progr.\ Theoret.\ Phys.\ Suppl.}
			{98}{198}{286}{1989}
			
\bibitem[JFU99]{JFU99}	\bibarticle{C.\ Jaff\'e, D.F.\ Farrelly, T.\ Uzer}
			{Transition state in atomic physics}
			{\PRA}{60}{3833}{3850}{1999}
			
\bibitem[JHY83]{Johnson/Hirschfelder/Yang}  
			\bibarticle{B.R.\ Johnson, J.D.\ Hirschfelder,
			K.H.\ Yang}
			{Interaction of atoms, molecules, and ions with
			constant electric and magnetic fields}
			{\RMP}{55}{109}{}{1983}

\bibitem[KvL95]{Kock/vanLeeuwen}  
			\bibarticle{P.M.\ Koch, K.A.H.\ van Leeuwen}
			{The importance of resonances in microwave
     			``ionization'' of excited hydrogen atoms}
			{Phys.\ Rep.} 
			{255}{289}{406}{1995}
		
\bibitem[L90]{Laskar1}
			\bibarticle{J.\ Laskar}	
			{The chaotic motion of the solar system: A numerical
			estimate of the size of the chaotic zones}
			{Icarus}{88}{266}{291}{1990} 
			
\bibitem[L96]{Laskar2}
			\bibarticle{J.\ Laskar}	
    			{Large scale chaos and marginal stability in the
			solar system}
			{Celestial Mech.\ Dynam.\ Astronom.}
			{64}{115}{162}{1996}
			
\bibitem[LR93]{LR93}	\bibarticle{J.\ Laskar, P.\ Robutel} 
			{The chaotic obliquity of the planets}
			{Nature}{361}{608}{612}{1993}
			
\bibitem[LL92]{Lichtenberg/Lieberman}  
			\bibbook{A.J.\ Lichtenberg, M.A.\ Lieberman}
			{Regular and Chaotic Dynamics}
			{Springer-Verlag}{New York, 1992}
			
\bibitem[Ma89]{Mathys}  	
			{G.\ Mathys},
			{}
			\bibref{Fundam.\ Cosm.\ Phys.}{13}{143}{}{1989}.
			
\bibitem[Mi82]{Mignard}  	
			{F.\ Mignard},
			{}
			\bibref{Icarus}{49}{347}{}{1982}.

\bibitem[MW92]{Main/Wunner2}  
			\bibarticle{J.\ Main, G.\ Wunner}
			{Ericson fluctuations in the chaotic ionization of
			the hydrogen atom in crossed magnetic and electric
			fields}
			{\PRL}{69}{586}{589}{1992}
			
\bibitem[N\&97]{NUFESWHD97}
			\bibarticle{C.\ Neumann {\it et al.}}
			{Symmetry breaking in crossed magnetic and electric
			fields}
			{\PRL}{78}{4705}{4708}{1997}

\bibitem[R63]{R63}	\bibarticle{P.J.\ Redmond} 
			{Generalization of the Runge-Lenz Vector in the
			Presence of an Electric Field}
			{\PR}{133}{B1352}{3}{1963}
			
\bibitem[RFW91]{Raithel/Fauth/Walther}  
			\bibarticle{G.\ Raithel, M.\ Fauth, H.\ Walther} 
			{Quasi-Landau resonances in the spectra of rubidium 
			Rydberg atoms in crossed electric and
			magnetic fields}
			{\PRA}{44}{1898}{1909}{1991}
			
\bibitem[RFW93]{Raithel/Fauth/Walther2}  
			\bibarticle{G.\ Raithel, M.\ Fauth, H.\ Walther}
			{Atoms in strong crossed electric and magnetic
			fields: Evidence for states with large
			electric-dipole moments}
			{\PRA}{47}{419}{440}{1993} 
		
\bibitem[RT97]{RT97}	\bibarticle{J.\ Rao, K.T.\ Taylor}
			{Atoms in crossed fields: calculations for barium
			and hydrogen}
                 	{J.\ Phys.\ B}{30}{3627}{3645}{1997}
		 
\bibitem[Schw16]{Schwarzschild}
			\bibtitle{K.\ Schwarzschild}
			{Sitzungsber.\ d.\ Berl.\ Akad.\ 1916},
			p.\ 548.
			
\bibitem[Schm93]{Schmelcher}  
			\bibarticle{P.\ Schmelcher}
			{Delocalization of excitons in a magnetic field}
			{\PRB}{48}{14642}{14645}{1993}

\bibitem[S83]{Solovev}  \bibarticle{E.A.\ Solov'ev}
			{Second-order perturbation theory for the hydrogen
			atom in crossed electric and magnetic fields}
			{Sov.\ Phys.\ JETP}{58}{63}{66}{1983}
			
\bibitem[TLL79]{TLL79}	\bibtitle{J.L.\ Tennyson, M.A.\ Lieberman, A.J.\
			Lichtenberg}
			{Diffusion in Near-Integrable Hamiltonian Systems
			with Three Degrees of Freedom}
			in \bibbook{M.\ Month, J.C.\ Herrera Eds.}
			{Nonlinear dynamics and the Beam--Beam Interaction}
			{Am.\ Inst.\ Phys.\ Conference Proceedings No.\ 57}
			{New York, 1979} pp.\ 272-301.
			
\bibitem[U\&91]{UDMRS91}
			\bibarticle{T.\ Uzer {\it et al.}}
			{Celestial Mechanics on a Microscopic Scale}
			{Science}{253}{42}{48}{1991}

\bibitem[UF95]{Uzer/Farrelly2}  
			\bibarticle{T.\ Uzer, D.\ Farrelly}
			{Threshold ionization dynamics of the hydrogen atom
			in crossed electric and magnetic fields}
			{\PRA}{52}{R2501}{R2504}{1995}

\bibitem[V96]{V96}	\bibbook{F.\ Verhulst}
			{Nonlinear Differential Equations and Dynamical
			Systems}
			{Springer-Verlag}
			{Berlin, 1996}

\bibitem[vMDU94]{vonMilczewski/Diercksen/Uzer}  
			\bibarticle{J.\ von\ Milczewski, G.H.F.\ Diercksen, 
			T.\ Uzer}
			{Intramanifold chaos in Rydberg atoms in external
			fields}
     			{\PRL}{73}{2428}{2431}{1994}

\bibitem[vMDU96]{vonMilczewski/Diercksen/Uzer3}	
			\bibarticle{J.\ von\ Milczewski, G.H.F.\ Diercksen, 
			T.\ Uzer} 
			{Computation of the Arnol'd Web for the hydrogen
			atom in crossed electric and magnetic fields}
			{\PRL}{76}{2890}{2893}{1996}
			
\bibitem[vMFU97a]{vMFU97}	
			\bibarticle{J.\ von\ Milczewski, D.\ Farelly, T.\ Uzer}
			{$1/r$ Dynamics in External Fields: 2D or 3D?}
			{\PRL}{78}{2349}{2352}{1997}
			
\bibitem[vMFU97b]{vMFU97b}
			\bibarticle{J.\ von\ Milczewski, D.\ Farelly, T.\ Uzer}
			{Role of the atomic Coulomb center in ionization and
			periodic orbit selection}
			{\PRA}{56}{657}{670}{1997}
			
\bibitem[vMU97a]{vonMilczewski/Diercksen/Uzer2}
			\bibarticle{J.\ von\ Milczewski, T.\ Uzer}
			{Chaos and order in crossed fields}
			{\PRE}{55}{6540}{6551}{1997}

\bibitem[vMU97b]{vMU97}	\bibarticle{J.\ von\ Milczewski, T.\ Uzer} 
			{Canonical perturbation treatment of a Rydberg
			electron in combined electric and magnetic fields}
			{\PRA}{56}{220}{231}{1997}
			
\bibitem[W\&89]{Wiebusch}	
			\bibarticle{G.\ Wiebusch {\it et al.}}
			{Hydrogen atom in crossed magnetic and electric
			fields}
			{\PRL}{62}{2821}{2824}{1989}

\bibitem[Y\&93]{Yeazell}
		  	\bibarticle{J.A.\ Yeazell {\it et al.}} 
			{Observation of wave packet motion along
			quasi-Landau orbits}
			{\PRL}{70}{2884}{2887}{1993}

\end{thebibliography}
\end{document}